\documentclass[a4paper,11pt,onecolumn,accepted=2023-10-17]{quantumarticle}

\pdfoutput = 1

\usepackage[utf8]{inputenc}
\usepackage[english]{babel}
\usepackage[T1]{fontenc}

\usepackage{lipsum}
\usepackage{mdframed}
\usepackage{booktabs}
\usepackage{longtable}

\usepackage[svgnames]{xcolor}
\usepackage[colorlinks=true,urlcolor = blue,linkcolor=blue,citecolor=teal,bookmarks=false]{hyperref}
\usepackage{amsfonts,amsmath,amssymb,bbm}
\usepackage{graphicx}
\usepackage{dsfont}
\usepackage{float}

\def\ii{{\rm i}}

\newcommand{\be}{\begin{equation} }
\newcommand{\ee}{\end{equation} }

\newcommand{\bp}{\begin{pmatrix}}
\newcommand{\ep}{\end{pmatrix}}

\usepackage[numbers, sort&compress]{natbib}

\begin{document}

\title{Integrable Quantum Circuits from the Star-Triangle Relation}

\author{Yuan Miao}
\affiliation{Kavli Institute for the Physics and Mathematics of the Universe (WPI),The University of Tokyo Institutes for Advanced Study, The University of Tokyo, Kashiwa, Chiba 277-8583, Japan}
\affiliation{Galileo Galilei Institute for Theoretical Physics, INFN, Largo Enrico Fermi, 2, 50125 Firenze, Italy}
\orcid{0000-0003-2086-1900}

\author{Eric Vernier}
\affiliation{Laboratoire de Probabilités, Statistique et Modélisation CNRS - Univ. Paris Cité - Sorbonne Univ. Paris, France}

\begin{abstract}

The star-triangle relation plays an important role in the realm of exactly solvable models, offering exact results for classical two-dimensional statistical mechanical models. In this article, we construct integrable quantum circuits using the star-triangle relation.  
Our construction relies on families of mutually commuting two-parameter transfer matrices for statistical mechanical models solved by the star-triangle relation, and differs from previously known constructions based on Yang-Baxter integrable vertex models. At special value of the spectral parameter, the transfer matrices are mapped into integrable quantum circuits, for which infinite families of local conserved charges can be derived. We demonstrate the construction by giving two examples of circuits acting on a chain of $Q-$state qudits: $Q$-state Potts circuits, whose integrability has been conjectured recently by Lotkov et al., and $\mathbb{Z}_Q$ circuits, which are novel to our knowledge. In the first example, we present for $Q=3$ a connection to the Zamolodchikov--Fateev 19-vertex model.
\end{abstract}

\maketitle

\section{Introduction}
\label{sec:intro}

Quantum circuits, built from a sequence of local operations acting on a system of qubits (or, more generally, qudits), have attracted an increasing interest over the past few years. First, they furnish a new playground for the investigation of many-body quantum physics, in particular for the study of out-of-equilibrium phenomena \cite{Nahum2017,Chan2018,Keyserlingk,Bertini2019,PiroliDualUnit}. Second, they can be implemented in a quantum computer and form the building blocks of digital quantum simulation \cite{Feynman_1982,Georgescu}. They can also be used to generate periodically-driven (Floquet) many-body systems, leading to exotic new phases of matter \cite{CayssolFloquet,KhemaniFloquet,ElseFloquet}. 

For many-body systems governed by continuous Hamiltonian evolution, the existence of integrable models has proven an invaluable tool in order to study physical properties both at equilibrium \cite{Bethe_1931, Baxter_1982, Gaudin_2014, Korepin_1993,takahashi_1999}, and out-of-equilibrium \cite{Calabrese_Essler_Mussardo_2016}. Quantum integrability usually refers to one-dimensional quantum Hamiltonians related to exactly solvable two-dimensional statistical mechanical models through the transfer matrix formalism and the Yang--Baxter equation, whose spectrum or correlation functions can typically be calculated exactly using tools such as the Bethe ansatz \cite{Gaudin_2014, Korepin_1993}. 
Beyond the possibility of exact results that it offers, integrability also comes with rich physical consequences. The existence of an extensive number of conserved quantities in integrable models constrains their late-time relaxation, yielding new equilibrium states known as Generalized Gibbs Ensembles \cite{Rigol_Dunjko_Olshanii_2008,Vidmar_Rigol_2016,Essler_Fagotti_2016}. For inhomogeneous systems integrability also constrains the transport properties, leading to Generalized Hydrodynamics \cite{CastroAlvaredo,Bertini_Collura}. 
It has therefore quickly become a natural question, whether one could similarly construct and study integrable models of quantum circuits, corresponding to dynamical models for one-dimensional quantum systems with discrete space and time.

It has long been known how to adapt the transfer matrix-mediated correspondence between integrable two-dimensional vertex models and quantum Hamiltonians to a circuit-like geometry \cite{Destri_1989, DDV, Faddeev_1994, Reshetikhin_1994}, in relation with the lattice regularisation of (1+1)-dimensional integrable quantum field theories. In the recent years this fact has been used to construct integrable Floquet dynamics \cite{Vanicat,Gritsev_2017,YM_Floquet, Yamazaki_2022}, and recently the effect of integrability on the late-time relaxation of digital quantum simulations has also been investigated \cite{VernierdGGE}. 
However, a systematic understanding of the condition when quantum circuits can be solved using quantum integrability is still missing. It is worth noting that most of the exact results obtained lately in fact concern quantum circuits which are solvable while escaping the traditional framework of Yang-Baxter integrability, namely, random  \cite{Nahum2017,Chan2018,Keyserlingk} and dual-unitary circuits \cite{Bertini2019,PiroliDualUnit}. There are also other examples on how to use quantum circuits to study quantum integrability that are different from our approach, see \cite{Nepomechie_2021, Economou_2021, Sopena_2022, Wilsmann_2018, Links_2022, Foerster_2022}.

In this work, we describe the construction of integrable quantum circuits based on $Q$-states spins with $\mathbb{Z}_Q$ symmetry. Those arise as generalizations of the Ising model (corresponding to $Q=2$), and can be realized with Rydberg atoms \cite{Samajdar_Choi_Pichler_Lukin_Sachdev_2018,Keesling}.
Furthermore, they have very rich physical properties, relating to quantum phase transitions and parafermions 
\cite{fradkin1980disorder,Fendley_2014,Alicea_Fendley_2016}.
Our construction uses a framework analogous to that of \cite{DDV}, namely inhomogeneous transfer matrices are used to generate a circuit-like dynamics, however in contrast with previous constructions the primary role for integrability is played here, rather than the Yang--Baxter equation, by the closely related Star-Triangle Relation (STR) \cite{Baxter_1982, AuYangPerk, Pokrovsky_1982}. 
Using known solutions of the star-triangle relation for $Q$-state spins, we construct two-parameter families of mutually commuting transfer matrices acting on a chain of $L$ spins.
At some special value of their parameters the transfer matrices become the generator of the circuit dynamics, while varying the parameters around their special value allows to construct local charges which are conserved by the dynamics.  

In practice, we focus in this work on two families of $Q$-states circuits, associated with two families of solutions of the STR: the so-called Potts circuits, whose integrability was conjectured in \cite{Gritsev_2022} (and for which the first few conserved charges were constructed by hand), and the so-called 
$\mathbb{Z}_Q$ circuits. 
The constructed circuits are in general interacting yet solvable, as guaranteed by the STR, and therefore go beyond some known results for driven Ising models that are solved using free fermionic techniques \cite{Oh_2013, Oh_2015, Teixeira_2020, Robinson_2021}. 
We would like to emphasize that, while most of this work is concerned with some particular $Q$-states models, our procedure works in principle for any solution of the Star-Triangle relation, and could be used to construct more generic integrable quantum circuits.

The paper is organized as follows. In Section \ref{sec:quantumcircuits}, we present some generic properties of the $Q$-states quantum circuits constructed in this work, and how they can be seen as emerging from the stroboscopic evolution of periodically driven (Floquet) systems. In Section \ref{sec:transfermat}, we present a generic procedure to construct quantum circuits from two-dimensional statistical mechanics model satisfying the Star-Triangle Relation. While this construction is not specific to $Q$-states systems and could in principle be applied more generically, in the rest of the paper we specify again to $Q$-states systems and construct two families of integrable quantum circuits. The first family, studied in Section \ref{sec:Potts}, is that of $Q$-states Potts circuits, where the $\mathbb{Z}_Q$ symmetry is enhanced to the symmetric group $S_Q$. We construct integrable circuits from previously known $S_Q$-symmetric solutions of the star-triangle relation \cite{Pokrovsky_1982}, and express the discrete time evolution operator as well as the conserved charges in terms of generators of the affine Temperley-Lieb algebra \cite{TL_1971}. The resulting dynamics is unitary, and can be thought of as the Floquet dynamics of a quantum Potts Hamiltonian. It recovers the circuit considered in \cite{Gritsev_2022}, and we also point out an interesting connection with the Zamolodchikov-Fateev 19-vertex model \cite{ZF_1980} and the Onsager algebra \cite{Onsager_1944}. 
The second family of models, which is the object of Section \ref{sec:AT}, is based on $\mathbb{Z}_Q$-symmetric solutions of the star-triangle relation \cite{Fateev_1982_2}. For $Q=3$, the resulting circuit coincides with the $S_3$-symmetric circuit of the first family. For general $Q>3$ however the constructed models differ from the previous ones, in particular they are not unitary. For $Q=4$, in particular, a relation is found with the critical Ashkin-Teller model \cite{Ashkin_Teller_1943, Alcaraz_1987, Alcaraz_1988}.

\section{$Q$-states quantum circuits}
\label{sec:quantumcircuits}

Before discussing the general framework for constructing integrable quantum circuits through the STR, which will be presented in Section \ref{sec:transfermat}, we start with a brief overview of the $Q$-states circuits which will be constructed from explicit solutions in Sections \ref{sec:Potts} and \ref{sec:AT}.  

One way to view those circuits is as stroboscopic (Floquet) evolution operators, motivated by the known results on periodically driven Ising models \cite{Oh_2013, Oh_2015, Teixeira_2020, Robinson_2021}. Such circuits were solved exactly by free fermionic techniques, and we consider in this work more generic cases which are intrinsically interacting.

We therefore consider a chain of $L$ consecutive $Q$-level spins (``qudits''), where $Q$ is some integer $\geq 2$. The total Hilbert space is the tensor product of the local $Q$-level spins, i.e. $(\mathbb{C}^Q)^{\otimes L}$. The quantum circuits that we study in this paper can be seen as a stroboscopic (Floquet) evolution of time-dependent quantum Hamiltonian $\mathbf{H} (t)$ such that
\be 
    \mathbf{H} ( t ) =  
	\begin{cases} \, \mathbf{H}_2 , \quad 0 \leq t < \tau , \\
		\, \mathbf{H}_1 , \quad \tau \leq t < 2 \tau ,
	\end{cases}
\ee 
which is periodic in time, i.e. $\mathbf{H} (t+2 n \tau ) = \mathbf{H} (t) $, $n \in \mathbb{Z}$. Furthermore, we assume that two parts $\mathbf{H}_1$ and $\mathbf{H}_2$ consist of terms acting on one or two consecutive sites of the $Q$-level spins respectively,
\be 
    \mathbf{H}_1 = \sum_{m = 1}^L \mathbf{h}^{(1)}_{m} , \quad \mathbf{H}_2 = \sum_{m = 1}^L \mathbf{h}^{(2)}_{m,m+1} .
\ee 

Periodic boundary condition is used here ($\mathbf{h}^{(2)}_{L,L+1} = \mathbf{h}^{(2)}_{L,1}$). We also assume that
\be 
    \left[ \mathbf{h}^{(1)}_{m} , \mathbf{h}^{(1)}_{n} \right] = 0 , \quad \left[ \mathbf{h}^{(2)}_{m,m+1} ,\mathbf{h}^{(2)}_{n,n+1} \right] = 0 , \quad \forall m,n .
    \label{eq:commuteassumption}
\ee 

\begin{figure}
    \centering
    \includegraphics[width=.82\linewidth]{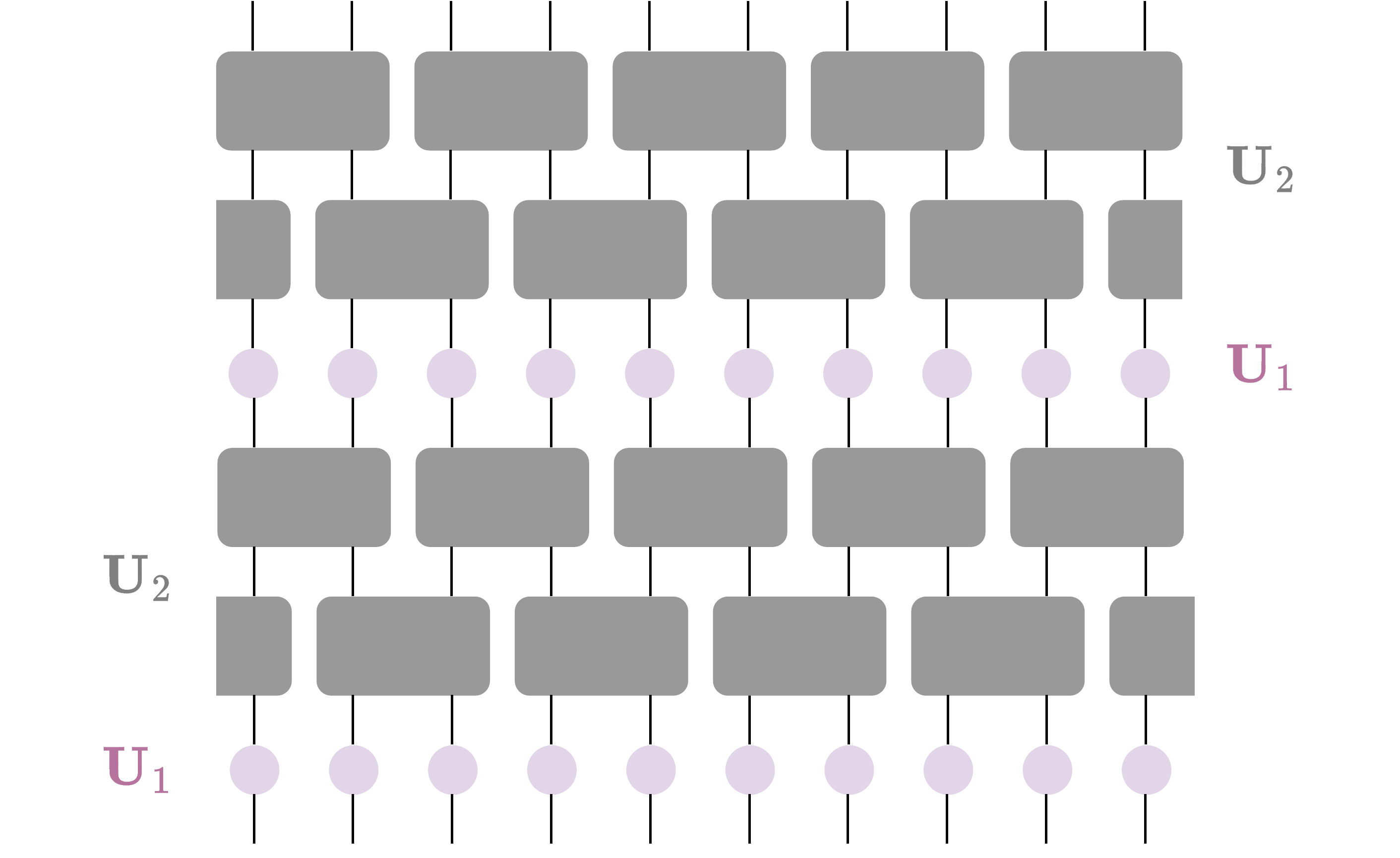}
    \caption{Generic structure of the circuits considered in this paper. The discrete time evolution is comprised of two steps, $\mathbf{U}_1$ which is the product of local one-site operations, and $\mathbf{U}_2$ which is the product of two-site gates. The two-site gates commute with one another and can be multiplied in arbitrary order. However, the two steps do not commute with each other, hence generating a non-trivial dynamics.}
    \label{fig:quantum_circuit_demo}
\end{figure}

In this case, the Floquet evolution operator $\mathbf{U}_{\rm F} (\tau) = \mathcal{P} \exp [\int_0^{2\tau} \mathrm{d} t \mathbf{H}(t) ]$, describing the stroboscopic time evolution of the time-dependent Hamiltonian $\mathbf{H}(t)$ \footnote{We can equivalently use a ``kicked'' time dependent Hamiltonian that gives the same stroboscopic time evolution. This will not change the quantum circuits that we study.}, becomes
\be 
    \mathbf{U}_{\rm F} (\tau ) = \exp \left( -\ii \mathbf{H}_1 \tau \right) \exp \left( -\ii \mathbf{H}_2 \tau \right) = \mathbf{U}_1 \mathbf{U}_2 ,
\ee 
Hence we can rewrite the stroboscopic time evolution $\mathbf{U}_{\rm F}^M (\tau)$, for an integer $M \in \mathbb{Z}_{>0}$, as a quantum circuit, as shown in Fig. \ref{fig:quantum_circuit_demo}. In particular, the stroboscopic time evolution of the kicked Ising model \cite{Gritsev_2017, Gritsev_2022, Robinson_2021} is of this type. 

Moreover, we would like to concentrate on models with a $\mathbb{Z}_Q$ ``clock'' symmetry, which generalizes the $\mathbb{Z}_2$ symmetry of the Ising model and connects with a number of interesting physical realizations \cite{Samajdar_Choi_Pichler_Lukin_Sachdev_2018,Keesling,fradkin1980disorder,Fendley_2014,Alicea_Fendley_2016}. For this sake we introduce the local operators $\mathbf{X}_m$, $\mathbf{Z}_m$ satisfying the following algebra
\be 
\label{eq:ZQalgebra}
	\mathbf{X}_m^\dagger = \mathbf{X}_m^{Q-1}, \mathbf{Z}_m^\dagger = \mathbf{Z}_m^{Q-1}  \quad \mathbf{X}_m^Q = \mathbf{Z}_m^Q = \mathbbm{1} ,  \quad \mathbf{X}_m \mathbf{Z}_m = \omega \mathbf{Z}_m \mathbf{X}_m   \,,
\ee 
where the $Q$-th root of unity $\omega = \exp \left( \frac{2 \ii \pi}{Q} \right)$, while operators acting on different spins commute :  $\mathbf{X}_m \mathbf{Z}_n = \mathbf{X}_n \mathbf{Z}_m$ for $m \neq n$ (see Eq. \eqref{XmZmdef} for an explicit representation). 
Requiring the assumption \eqref{eq:commuteassumption}, we focus on the cases where the Floquet evolution operator $\mathbf{U}_{\rm F} = \mathbf{U}_1 \mathbf{U}_2$ is decomposed as
\be 
\begin{split}
\mathbf{U}_1 & = \prod_{j=1}^L \left( \sum_{a=1}^{Q-1} u_a (\mathbf{X}_{j})^a \right) \\
\mathbf{U}_2 & = \prod_{j=1}^L \left( \sum_{a=1}^{Q-1} v_a (\mathbf{Z}_{j}^\dag \mathbf{Z}_{j+1})^a\right) \,.
\end{split}
\label{eq:ZQFloquet}
\ee 

Written in the above form, the evolution generators $\mathbf{U}_1$ and $\mathbf{U}_2$ are manifestly $\mathbb{Z}_Q$-symmetric, namely invariant under the operation $\mathbf{Z}_j \to \omega \mathbf{Z}_j$, $\mathbf{X}_j \to \mathbf{X}_j$ applied simultaneously on all spins. 
Moreover, in all examples considered in the following they will turn out to enjoy another symmetry encoded in the fact that $u_{Q-a}=u_a$ and $v_{Q-a}=v_a$ for all $a$, namely they are invariant under the charge conjugation operation $\mathbf{Z}_j \leftrightarrow \mathbf{Z}_j^\dagger$, $\mathbf{X}_j \leftrightarrow \mathbf{X}_j^\dagger$. 
For $Q=3$, the $\mathbb{Z}_3$ symmetry and charge conjugation together generate a $S_3$ symmetry group. For $Q\geq 4$ the $\mathbb{Z}_Q$ (+ charge conjugation) and $S_Q$ symmetries cease to be equivalent, and we will consider both types of models, invariant under the $S_Q$ and $\mathbb{Z}_Q$ symmetry respectively.

{\bf Remarks.} For generic choices of the parameters $u_a$ and $v_a$, the resulting quantum circuits are not integrable (or exactly solvable). As we shall explain in the latter sections, certain choices of the parameters $u_a$ and $v_a$ will lead to the integrable quantum circuits that commute with transfer matrices. One notable example is when $u_a = v_b$ for arbitrary $a , b \in \mathbb{Z}_Q$, which has been conjectured in \cite{Gritsev_2022}. We shall prove the conjecture in Sec. \ref{sec:Potts} and provide a different example in Sec. \ref{sec:AT}. 
Another crucial remark is about the unitarity of the Floquet evolution operator $\mathbf{U}_{\rm F}$ (or subsequently the operators $\mathbf{U}_1$ and $\mathbf{U}_2$). In fact, arbitrary choices of the parameters $u_a$ and $v_a$ will not lead to a unitary time evolution. An exception occurs with the Potts circuits explained in Sec. \ref{sec:Potts}, cf. \eqref{eq:Pottscircuits1}.

\section{Two-parameter transfer matrices from the Star-Triangle Relation}
\label{sec:transfermat}

\subsection{The Star-triangle relation}
\label{sec:STR}

The star-triangle relation  (STR) \cite{Baxter_1982, AuYangPerk, Pokrovsky_1982} is a powerful tool to solve 2-dimensional statistical mechanical models exactly. Several renowned statistical mechanical models can be solved by the STR, such as classical Ising model, classical (chiral) Potts models on a square lattice, etc... 

\begin{figure}
\centering
\includegraphics[width=.75\linewidth]{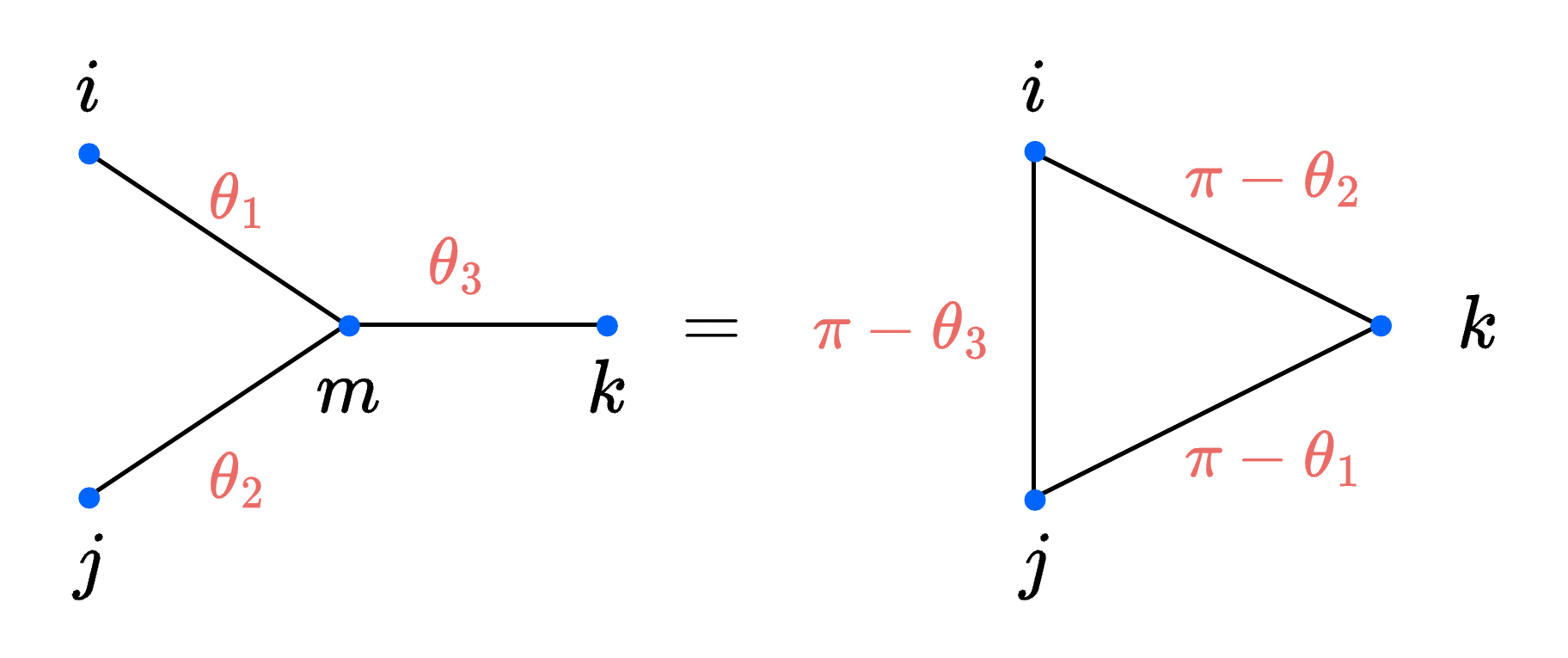}
\caption{Graphical illustration of the star-triangle relation \eqref{eq:STR}.}
\label{fig:star_triangle_demo}
\end{figure}

Generically, the star-triangle relation is defined for a statistical model of ``heights'', or ``spins'' taking values in some set $\mathcal{S}\subset \mathbb{Z}$. For the moment we do not need to specify further the nature of $\mathcal{S}$, but turning to explicit solutions of the star-triangle relation in Sections \ref{sec:Potts} and \ref{sec:AT}, it will taken to be $\{1,\ldots Q\}$, with $Q$ some positive integer (in other terms the heights are defined modulo $Q$). The heights sit at the vertices of a two-dimensional lattice and the weight of a given height configuration is the product over all edges of a function $K (\theta ; i , j )$ of the adjacent heights $i, j$ , where $\theta \in \mathbb{C}$ is an additional parameter called spectral parameter. 
The star-triangle relation then reads \cite{Baxter_1982} 
\be 
\begin{split}
  \sum_{m \in \mathcal{S}} K(\theta_1 ; i , m ) & K(\theta_2 ; j , m ) K(\theta_3 ; k , m ) \\
  & = f(\theta_1 , \theta_2, \theta_3 ) K( \pi - \theta_1 ; j , k ) K(\pi - \theta_2 ; k , i ) K(\pi - \theta_3 ; i , j ) , \\
  \theta_1 + \theta_2 + \theta_3 & = \pi \,, 
\end{split}
\label{eq:STR}
\ee 
where $f(\theta_1,\theta_2,\theta_3)$ is some normalization function which does not depend on the heights $i,j,k$.
A pictorial illustration of \eqref{eq:STR} is given in Fig. \ref{fig:star_triangle_demo}.

In the following we will assume that the function $K (\theta ; i , j )$ satisfies the following additional properties:
\begin{align} 
  K(\theta ;  i , j ) &= K(\theta ;  i - j ) = K(\theta ;  j - i )   \label{eq:Kdifference} \,, \end{align} 
While there exist solutions of the star-triangle relation which do not verify Eq. \eqref{eq:Kdifference} the latter is verified in many cases of physical relevance, and will be in particular for the solutions of considered in this work.     
Furthermore, all solutions of the star-triangle relation considered in this work allow for two special values of the spectral parameter, $\theta=0,\pi$, for which the function $K(\theta,\alpha,\beta)$ takes a particularly simple form :
\be
      K(0 ;  i , j ) = \delta_{i,j } \,, 
      \qquad      
            K(\pi ;  i , j ) = \kappa \, , \qquad  \forall i , j \in \mathcal{S} \,,
\label{eq:shiftpoints}
\ee 
where the parameter $\kappa$ entering the second equation is independent of the indices $\alpha,\beta$.

\subsection{Two-parameter transfer matrices}

\begin{figure}
\centering
\includegraphics[width=.8\linewidth]{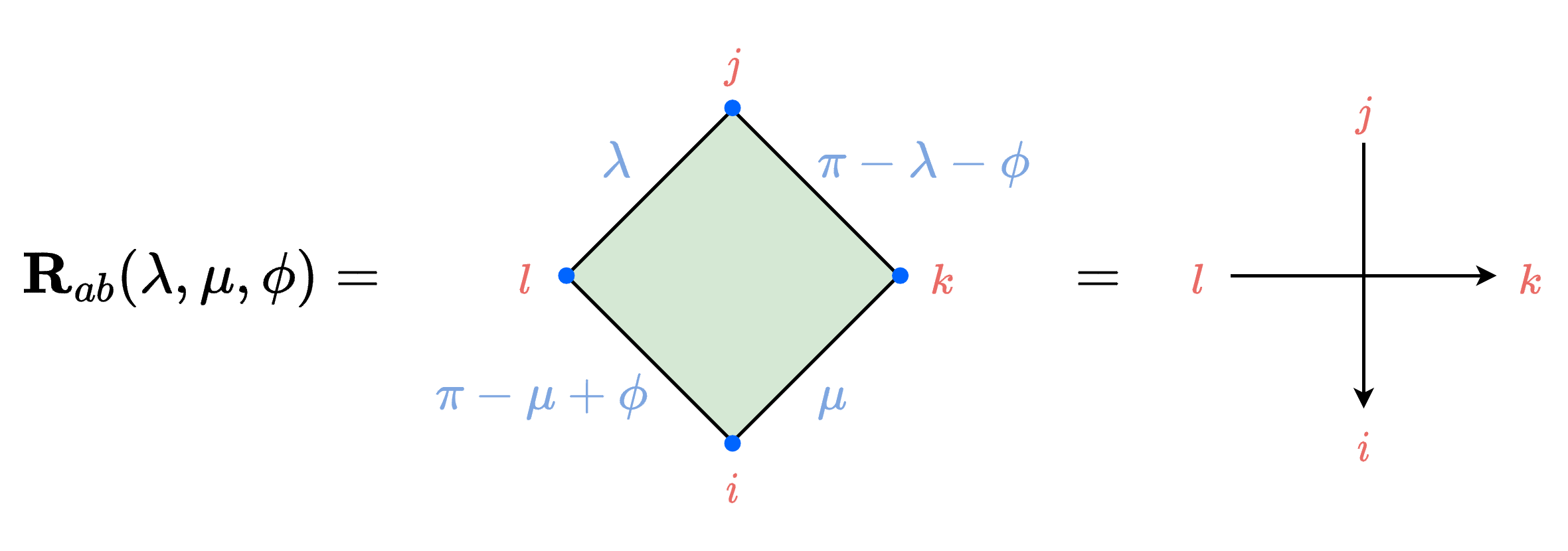}
\caption{Pictorial illustration of the R matrix of Eq. \eqref{eq:Rmatdef}.}
\label{fig:R_mat}
\end{figure}

From the star-triangle relation \eqref{eq:STR}, we can construct a set of mutually commuting transfer matrices, which can conveniently be recast as the row-to-row transfer matrices of a vertex model. To achieve this, we follow the route of  \cite{baxter1988new}. 
We start by grouping the interactions along the edges surrounding a given ``plaquette'' into the following R matrix (see Figure \ref{fig:R_mat})
\be 
  \mathbf{R}_{a b} (\lambda , \mu , \phi ) = \sum_{i,j,k,l \in \mathcal{S}} K(\lambda ; l , j ) K(\pi - \lambda - \phi ; j , k ) K(\mu ; i , k ) K(\pi - \mu + \phi ; l , i )   \mathbf{E}_a^{i,j} \otimes \mathbf{E}_b^{k,l} \,,
  \label{eq:Rmatdef}
\ee
where the Kronecker matrices $\mathbf{E}_a^{i,j}$, $\mathbf{E}_b^{k,l}$ act in vector spaces $a$ and $b$ whose basis states are indexed by the states in  $\mathcal{S}$.  

\begin{figure}
\centering
\includegraphics[width=.75\linewidth]{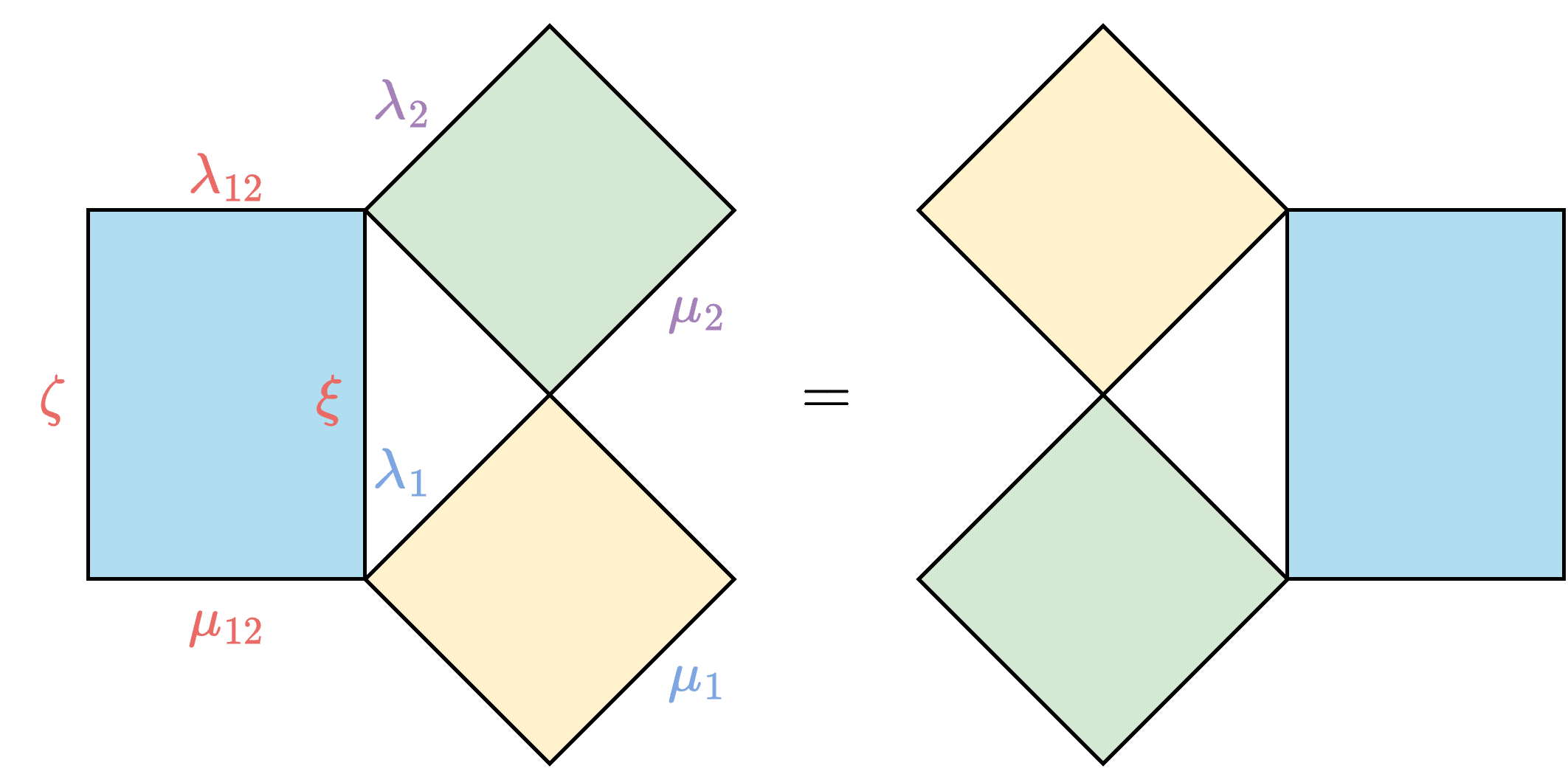}
\caption{A diagrammatic description of the Yang--Baxter relation in \eqref{eq:YBE1}. The spectral parameters are $\xi = \pi - \lambda_1 + \mu_2 - \phi$, and $\zeta = \pi - \mu_1 + \lambda_2 + \phi$.}
\label{fig:YBE1}
\end{figure}

As detailed in App. \ref{app:YBE}, it can be shown using the star-triangle relation that the R matrix obeys the Yang--Baxter equation 
\be
 \mathbf{R}_{ab} (\lambda_{12} , \mu_{12}, \phi^\prime ) \mathbf{R}_{ac} (\lambda_1 , \mu_1 , \phi ) \mathbf{R}_{bc} (\lambda_2 , \mu_2 , \phi ) 
 = \mathbf{R}_{bc} (\lambda_2 , \mu_2 , \phi ) \mathbf{R}_{ac} (\lambda_1 , \mu_1 , \phi )   \mathbf{R}_{ab} (\lambda_{12} , \mu_{12}, \phi^\prime )  \,,
\label{eq:YBE1}
\ee
where
\be 
  \lambda_{12} = \lambda_1 - \lambda_2 , \quad \mu_{12} = \mu_1 - \mu_2 , \quad \phi^\prime = \phi + \lambda_1 - \mu_1 .
\ee
The pictorial interpretation of the Yang--Baxter equation in terms of plaquettes is given in Fig. \ref{fig:YBE1}.

Using the R matrix, we can group the weights of all plaquettes along a horizontal row of the rotated square lattice into the following matrix product operator called transfer matrix  
\be 
  \mathbf{T} (\lambda , \mu , \phi , \{ \zeta_j \} ) = \mathbf{Tr}_a \left[ \prod_{j=1}^L \mathbf{R}_{a j} (\lambda - \zeta_j , \mu - \zeta_j , \phi ) \right] ,
\label{eq:Tmatdef}
\ee 
where the trace $\mathbf{Tr}_a$ follows from the choice of periodic boundary conditions in the horizontal direction, and where $\{ \zeta_j \}$ are arbitrary spectral parameters, which can generically taken to be inhomogeneous.  In the literature, people usually consider the case with $\phi = 0$ and inhomogeneities $\zeta_j = 0$, which has been used as the transfer matrix for quantum Potts chain or clock Hamiltonians \cite{baxter1988new}. In contrast, in the present case, we will need the parameter $\phi \neq 0$ to establish a connection with integrable quantum circuits. 
The transfer matrix is depicted pictorially in  Fig. \ref{fig:transfermatdecomp}, where our convention is that it transfer the heights of the top row to the bottom row.

From the Yang--Baxter equation \eqref{eq:YBE1}, it can be shown that the transfer matrices with the same $\phi$ and inhomogeneities $\{\zeta_j\}$ but different horizontal spectral parameters $\lambda,\mu$ commute : 
\be 
  \left[ \mathbf{T} (\lambda_1 , \mu_1 , \phi , \{ \zeta_j \} ) , \mathbf{T} (\lambda_2 , \mu_2 , \phi , \{ \zeta_j \} )  \right] = 0 , \quad \lambda_1 , \lambda_2 , \mu_1 , \mu_2 \in \mathbb{C} \,.
\ee 
Therefore, we will often call these ``two-parameter transfer matrices'', meaning that for a given model $\phi$ and $\{\xi_j\}$ are fixed while $\lambda$ and $\mu$ are allowed to vary.
In the remaining part of the article, we will focus on the homogeneous case where all the $\zeta_j \to 0$, and will therefore omit the latter from our notations.

From the star-triangle relation \eqref{eq:STR}, the two-parameter transfer matrix satisfies a ``self-duality'' relation, i.e.
\be 
  \mathbf{T} (\lambda , \mu , \phi ) = \mathbf{T} (\mu - \phi , \lambda + \phi , \phi ) .
  \label{eq:selfdual}
\ee
A diagrammatic derivation of the self-dual relation is demonstrated in Fig. \ref{fig:selfdual} in App. \ref{app:selfdual}.

In addition, considering the product of two transfer matrices, and applying the star-triangle relation \eqref{eq:STR}, we have
\be 
\mathbf{T} (\lambda_1 , \mu_1 , \phi ) \mathbf{T} (\lambda_2 , \mu_2 , \phi ) = \mathbf{T} (\mu_2 - \phi , \mu_1 , \phi ) \mathbf{T} (\lambda_1 , \lambda_2 + \phi , \phi ) .
\label{eq:factor1}
\ee 
The proof is analogous to the ``self-dual'' property and the diagrammatic demonstration is shown in Fig. \ref{fig:factor1} in App. \ref{app:factorder}.

Combining with the ``self-duality'' of the transfer matrix \eqref{eq:selfdual}, we show the factorisation of the two-parameter transfer matrix,
\be 
\begin{split}
\mathbf{T} (\lambda_1 , \mu_1 , \phi ) \mathbf{T} (\lambda_2 , \mu_2 , \phi ) &= \mathbf{T} (\mu_2 - \phi , \mu_1 , \phi ) \mathbf{T} (\lambda_1 , \lambda_2 + \phi , \phi ) \\
& = \mathbf{T} (\mu_1 - \phi , \mu_2 , \phi ) \mathbf{T} (\lambda_2 , \lambda_1 + \phi , \phi ) \\
& = \mathbf{T} (\lambda_1 , \mu_2 , \phi ) \mathbf{T} (\lambda_2 , \mu_1 , \phi ) .
\end{split} 
\label{eq:factorisation}
\ee

Therefore, we define two operators $\mathbf{Q} (\lambda)$ and $\mathbf{P} (\mu)$, such that
\be 
    \mathbf{Q} (\lambda ) = \mathbf{T} (\lambda , 0,\phi), \quad \mathbf{P} (\mu ) = \mathbf{T} (0, \mu,\phi) \mathbf{T}^{-1} (0,0,\phi) .
\ee 
We have assumed that $\mathbf{T} (0,0)$ is invertible, which is the case for the examples below. The two operators commute, i.e.
\be 
    [\mathbf{Q} (\lambda ) , \mathbf{Q} (\mu )] = [\mathbf{P} (\lambda ) , \mathbf{P} (\mu )]  = [\mathbf{Q} (\lambda ) , \mathbf{P} (\mu )] = 0 , \quad \forall \lambda , \mu \in \mathbb{C} .
\ee

In this way, the two-parameter transfer matrix is factorised into two parts,
\be 
    \mathbf{T} (\lambda , \mu,\phi) = \mathbf{Q} (\lambda) \mathbf{P} (\mu) ,
    \label{eq:TQP}
\ee
by using the factorisation property \eqref{eq:factorisation}.

In the meantime, the self-duality implies
\be 
    \mathbf{T} (\lambda , \mu,\phi) = \mathbf{Q} (\mu - \phi ) \mathbf{P} (\lambda + \phi ) .
\ee 
We notice the resemblance to the two-parameter transfer matrix of the 6-vertex model at root of unity, which can be used to construct Baxter's Q operator \cite{YM_spectrum_XXZ}.

\subsection{Derivation of local commuting charges}

When the function $K(\theta ; i , j )$ satisfies 
\be \label{eq:K0KK}
\begin{split}
& K( 0 ; i , j) = \delta_{i, j} , 
\\
& K (\pi - \phi ; i , j ) K (\pi + \phi ; i , j ) = f (\phi) , \quad \forall i, j \in \mathcal{S} ,
 \end{split}
\ee
as in the case of all examples considered below in Sections \ref{sec:Potts} and \ref{sec:AT}, we have
\be 
  \mathbf{R}_{a,b} (0,0,\phi ) = f (\phi) \mathbf{P}_{a,b} ,
\ee
where the operator $\mathbf{P}_{a,b}$ is the permutation operator such that $\mathbf{P}_{a,b} \mathbf{O}_a \mathbf{P}_{a,b} = \mathbf{O}_b$.

In this case, the two-parameter transfer matrix becomes
\be 
  \mathbf{T} (0,0,\phi ) = \mathrm{Tr}_a \left( \prod_{j=1}^L \mathbf{P}_{a,j} \right) = \prod_{j=L-1}^1 \mathbf{P}_{j,j+1} = \mathbf{G}^{-1} ,
\ee
where the operator $\mathbf{G} = \prod_{j=1}^{L-1} \mathbf{P}_{j,j+1}$ is the one site translation operator.

In this scenario,
\be 
    \mathbf{Q} (0) = \mathbf{G}^{-1} , \quad \mathbf{P} (0) = \mathbbm{1}.
\ee 
and a family of mutually commuting local conserved charges can be constructed by taking the logarithmic derivatives of the transfer matrix around the point $\lambda=0$, $\mu=0$,
\be 
    \mathbf{I}_{m,n} = \left. \partial_\lambda^m \partial_\mu^n \log \mathbf{T} (\lambda , \mu ,\phi) \right|_{\lambda = 0 , \mu = 0} , \quad m,n \in \mathbb{Z}_{>0} .
\ee
Due to the factorised form of the two-parameter transfer matrix \eqref{eq:TQP}, we have
\be 
    \mathbf{I}_{m,n} = 0 , \quad m \neq 0 , \, n \neq 0 .
\ee
There are therefore two sets of independent conserved quantities (when $\phi \neq 0$), namely
\be 
    \mathbf{I}_{m,0} , \,\, \mathbf{I}_{0,n} , \quad m,n \in \mathbb{Z}_{>0} .  
\ee 
Note that when $\phi = 0$, $\mathbf{I}_{m,0} = \mathbf{I}_{0,m}$.

\subsection{Circuit geometry}

\begin{figure}
    \centering
    \includegraphics[width=.9\linewidth]{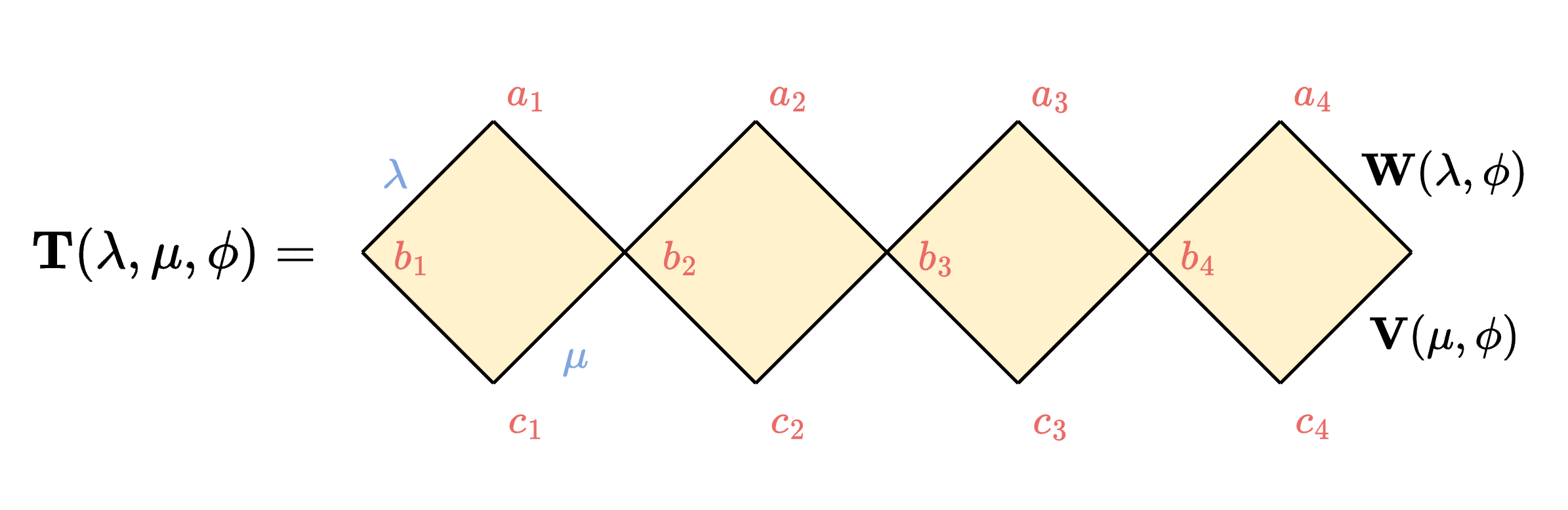}
    \caption{The two-parameter transfer matrix $T(\lambda,\mu,\phi)$, transfering the heights of the top row ($a_1,a_2,\ldots$) to the bottom row ($c_1,c_2,\ldots$).}
    \label{fig:transfermatdecomp}
\end{figure}

In order to recover a circuit-like geometry, we introduce another way of decomposing the two-parameter transfer matrix,
\be 
    \mathbf{T} (\lambda , \mu , \phi ) = \mathbf{V} (\mu , \phi) \mathbf{W} (\lambda,\phi) \,,
    \label{eq:transfermatdecomp}
\ee
where the matrices $\mathbf{V} (\mu , \phi)$ and $\mathbf{W} (\lambda,\phi) $ encode the weights of the two lower (resp. upper) edges of each plaquette, as illustrated in Fig. \ref{fig:transfermatdecomp}.
More precisely, they have the following matrix elements
\begin{align} 
    \mathbf{V}^{c_1 , c_2, \cdots c_L}_{b_1, b_2 , \cdots b_L} (\mu, \phi )
    &=  K(\mu ; c_1,b_2)K(\pi-\mu+\phi ; c_2,b_2)
    \ldots 
K(\mu; c_L,b_1)K(\pi-\mu+\phi ; c_1,b_1) \,,
\\
    \mathbf{W}^{b_1 , b_2, \cdots b_L}_{a_1, a_2 , \cdots a_L} (\lambda, \phi ) 
    &= K(\pi-\lambda-\phi; b_2,a_1)K(\lambda; b_2,a_2)
    \ldots 
    K(\pi-\lambda-\phi; b_{1},a_L)K(\lambda; b_1,a_1) 
    \,.
\end{align} 

This decomposition is different from the factorisation \eqref{eq:factorisation}, in particular
\be 
    [\mathbf{W}(\lambda_1 , \phi ) , \mathbf{W}(\lambda_2 , \phi ) ] \neq 0 , \quad [\mathbf{V}(\lambda_1 , \phi ) , \mathbf{V}(\lambda_2 , \phi ) ] \neq 0 ,
\ee
for generic $\lambda_1$, $\lambda_2$.

\begin{figure}
\centering
\includegraphics[width=.94\linewidth]{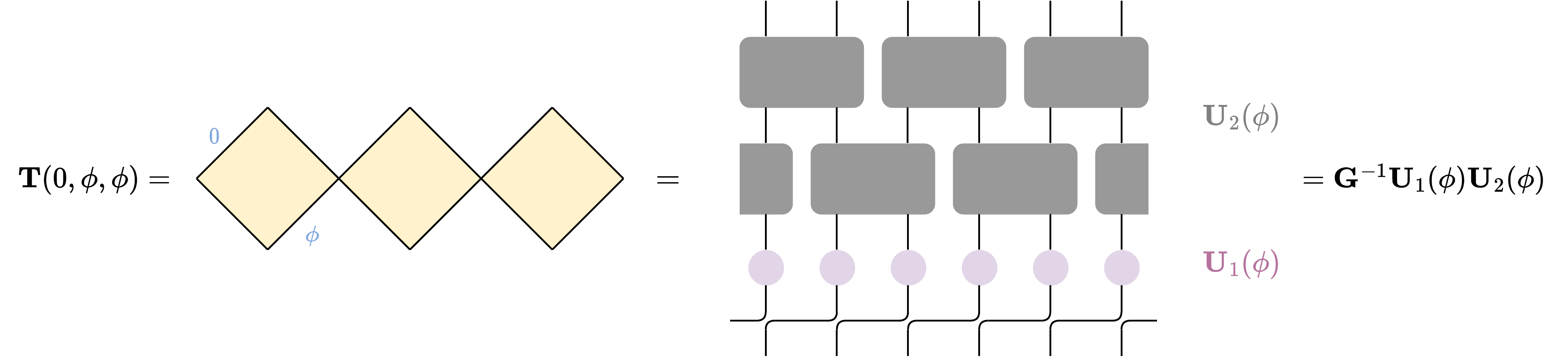}
\caption{Relation between the two-parameter transfer matrix and the integrable quantum circuit. 
}
\label{fig:transfermatcircuit}
\end{figure}

Let us now specify the spectral parameters to $\lambda,\mu = 0, \phi$. 
In this case, using the special values 
\eqref{eq:shiftpoints} of the function $K(\ldots,a,b)$, we find : 
\begin{align} 
    \mathbf{V}^{c_1 , c_2, \cdots c_L}_{b_1, b_2 , \cdots b_L} (\phi, \phi )
    &= \kappa^L K(\phi;c_1,b_2) \ldots K(\phi;c_L,b_1) \,,
\\
    \mathbf{W}^{b_1 , b_2, \cdots b_L}_{a_1, a_2 , \cdots a_L} (0, \phi ) 
    &= \delta_{b_1,a_1} \ldots \delta_{b_L,a_L} K(\pi-\phi;a_2,a_1) \ldots K(\pi-\phi;a_{1},a_L) 
    \,.
\end{align}
We can therefore rewrite : 
\be 
    \mathbf{V} (\phi , \phi) =  \mathbf{G}^{-1} \mathbf{U}_1 (\phi ) , \quad \mathbf{W} (0, \phi) =  \mathbf{U}_2 (\phi ) ,
\ee
where $\mathbf{G}^{-1}$ is the inverse translation operator introduced in the previous section, and $\mathbf{U}_1 (\phi )$ and $\mathbf{U}_2 (\phi )$ are products of single-site operators and double-site operators, respectively.  
The transfer matrix can therefore be expressed as the generator of a discrete quantum circuit dynamics,
\be 
  \mathbf{T} (0,\phi,\phi ) = \mathbf{G}^{-1} \mathbf{U}_1 (\phi ) \mathbf{U}_2 (\phi )  ,
\ee
with
\be 
    \left[\mathbf{G}^{-1}, \mathbf{U}_1 (\phi ) \right] = \left[ \mathbf{G}^{-1} , \mathbf{U}_2 (\phi ) \right] = 0 ,
\ee
as shown in Fig. \ref{fig:transfermatcircuit}. 

Defining the discrete time evolution operator 
\be 
\mathbf{U}_{\rm F} (\phi ) = \mathbf{U}_1 (\phi ) \mathbf{U}_2 (\phi )   =  \mathbf{G}  \mathbf{T} (0, \phi , \phi ) \,, 
\ee 
and using the fact that $\left[ \mathbf{G} , \mathbf{T} (\lambda , \mu , \phi) \right] = 0 $ for all $\lambda,\mu$, we therefore see that $\mathbf{U}_{\rm F}(\phi)$ commutes with the two-parameter family of transfer matrices,
\be 
 \left[ \mathbf{U}_{\rm F} (\phi) , \mathbf{T} (\lambda , \mu , \phi) \right] = 0 , \,\, \lambda , \mu \in \mathbb{C} ,
 \label{eq:commutationwithT}
\ee
and therefore with the charges $\mathbf{I}_{m,0}$ and $\mathbf{I}_{0,n}$ constructed in the previous section. In this sense, it defines an integrable discrete dynamics. 
In the following two sections we will demonstrate this construction using known families of solutions of the star-triangle relation, associated respectively with the $Q$-state Potts model and the Fateev--Zamolodchikov $\mathbb{Z}_Q$ model.

{\bf Remark.} Alternatively, if we set the inhomogeneities $\{ \zeta_j \}$ to be staggered 
\be 
\zeta_{2m-1} = \zeta_1 , \quad \zeta_{2m} = \zeta_2 , \quad \forall m \in \mathbb{Z}_+ ,
\ee
we can construct a different integrable quantum circuits with brick-wall structure, cf. Fig. 3 of \cite{YM_Floquet}, via the ``Floquet Baxterisation'' \cite{YM_Floquet}. The procedure is described in details in Sec. 4 of \cite{YM_Floquet}.

\section{Example: $Q$-state Potts circuits}
\label{sec:Potts}

We now move on to $Q$-states model, with $Q$ some positive integer. 
Namely, we now specify the generic exposition of Section \ref{sec:transfermat} to statistical models where the set of allowed heights at each site is $\mathcal{S}=\{1,\ldots Q\}$, and will derive from there quantum circuits of the form discussed in Section \ref{sec:quantumcircuits}. 
In this Section we focus on one of the most renowned examples, that of the $Q$-state Potts model \cite{Baxter_1982}. To begin with, we define the parameter $\eta$ as
\be 
  \sqrt{Q} = 2 \cosh \eta . 
\ee 

A qualitative feature separates the regimes $Q\leq 4$ (for which the Potts model has a second-order phase transition) and $Q>4$ (for which this transition becomes first order), which can be seen at the level of the parameter $\eta$\cite{Baxter_1982} For $Q=2,3$ it is pure imaginary, namely $\eta = \frac{\ii \pi}{4}$ and $\eta = \frac{\ii \pi}{6}$ respectively, while for $Q\geq 5$, $\eta \in \mathbb{R}$. The $Q=4$ case corresponding to $\eta=0$ is special, as in this case the star-triangle relation becomes rational instead of trigonometric (cf. \eqref{eq:STRPotts4}).

\subsection{Star-Triangle relation and two-parameter R matrix}

The specificity of the Potts model is that it is invariant under the permutation group $S_Q$ of internal indices, and a corresponding solution of the star-triangle equation has been found under the form \cite{Baxter_1982, Pokrovsky_1982}
\be 
K_{\rm Potts}(\theta ; a , b) = \frac{1}{\sqrt{Q} \sin (\eta/\ii)} \sin \left( \frac{\eta \theta}{\ii \pi} \right) + \frac{1}{\sin (\eta/\ii) } \sin \left( \frac{\eta (\pi - \theta) }{\ii \pi} \right) \delta_{a,b} .
\label{eq:STRPotts}
\ee

To be specific, we write down the explicit expressions of the solution \eqref{eq:STRPotts} with $Q = 2,3,4$,
\be 
    K_{\rm Potts}(\theta ; a , b) = \sin \left( \frac{\theta}{4} \right) + \sqrt{2} \sin \left( \frac{\pi - \theta}{4} \right) \delta_{a,b} , \quad Q = 2 ;
\ee
\be 
    K_{\rm Potts}(\theta ; a , b) = \frac{2}{\sqrt{3}} \sin \left( \frac{\theta}{6} \right) + 2 \sin \left( \frac{\pi - \theta}{6} \right) \delta_{a,b} , \quad Q = 3 ;
\ee
\be 
  K_{\rm Potts}(\theta ; a , b) = \frac{\theta}{2 \pi} + \left( 1 - \frac{\theta}{\pi} \right) \delta_{a,b} , \quad Q = 4 .
\label{eq:STRPotts4}
\ee

Using the star-triangle relation, we construct the two-parameter R matrix in the manner of \eqref{eq:Rmatdef}, satisfying the Yang-Baxter relation \eqref{eq:YBE1}. The two-parameter transfer matrix can be constructed using \eqref{eq:Tmatdef}.
The solution \eqref{eq:STRPotts} satisfies properties of the form \eqref{eq:K0KK}, where 
\be
f(\phi) = \begin{cases}
\frac{4}{Q(4-Q)} \sin \left( \frac{\eta (\pi-\phi)}{\ii \pi} \right) \sin \left( \frac{\eta (\pi+\phi)}{\ii \pi}\right) \qquad \text{for}~Q\neq 4 
\\
\frac{(\pi-\phi)(\pi+\phi)}{4\pi^2} \qquad \text{for}~Q=4
\end{cases} 
\ee 
Therefore, the R matrix \eqref{eq:Rmatdef} satisfies
\be 
    \mathbf{R}_{a,b} (0,0, \phi) = f(\phi)  \mathbf{P}_{a,b} .
\label{eq:Rmatnormalisation} 
\ee 
When $Q=3$, the normalisation factor becomes
\be 
    \mathbf{R}_{a,b} (0,0, \phi) = \frac{2 \cos (\phi/3)-1}{3} \mathbf{P}_{a,b} ,
\ee
which we will focus on later.

{\bf Remark.} When parameter $\phi = 0$, the two-parameter transfer matrix becomes the transfer matrix of the 3-state Potts model \cite{Baxter_1982, Alcaraz_1986, Albertini_1992_1}.

\subsection{Quantum circuit}
\label{sec:Potts/circuit} 

As anticipated in Section \ref{sec:quantumcircuits}, a convenient way to express the circuit operators obtained from the two-parameter transfer matrices is to introduce the Potts operators acting on the physical Hilbert space $\left( \mathbb{C}^Q \right)^{\otimes L}$,
\be 
\begin{split}
	\mathbf{X}_m = \mathbbm{1}^{\otimes (m-1)} \otimes \left( \mathbf{E}^{Q,1}_m + \sum_{j=1}^{Q-1} \mathbf{E}^{j,j+1}_m \right) \otimes \mathbbm{1}^{\otimes (L-m)} , \\
	\mathbf{Z}_m = \mathbbm{1}^{\otimes (m-1)} \otimes \left(  \sum_{j=1}^{Q} \omega^{j-1} \mathbf{E}^{j,j}_m \right) \otimes \mathbbm{1}^{\otimes (L-m)} , 
 \label{XmZmdef}
\end{split}
\ee 
where the $Q$-th root of unity $\omega = \exp \left( \frac{2 \ii \pi}{Q} \right)$. Those can be easily checked to satisfy the algebra \eqref{eq:ZQalgebra}.

Another sets of useful operators are the Potts representation of the affine Temperley--Lieb algebra \cite{TL_1971, Baxter_1982, YM_Onsager_2},
\be 
  \mathbf{e}_{2m-1} = \frac{1}{\sqrt{Q}} \sum_{a=0}^{Q-1} \mathbf{X}^a_m, \quad \mathbf{e}_{2m} = \frac{1}{\sqrt{Q}} \sum_{a=0}^{Q-1} \left( \mathbf{Z}^{\dag}_m \mathbf{Z}_{m+1} \right)^a ,
 \label{eq:aTLdef}
\ee
which satisfy the following relations,
\be 
	\mathbf{e}_{m}^2 = \sqrt{Q} \mathbf{e}_m , \quad \mathbf{e}_{m} \mathbf{e}_{m\pm 1} \mathbf{e}_{m} = \mathbf{e}_m , \quad \mathbf{e}_{m} \mathbf{e}_{n} = \mathbf{e}_{n} \mathbf{e}_{m} , \,\, |m-n| \geq 2 ,
\ee 
with periodic boundary condition $\mathbf{e}_{2L+1} = \mathbf{e}_{1}$. Furthermore, these are manifestly hermitian, $\mathbf{e}_m^\dagger = \mathbf{e}_m$.

Following the circuit construction of Section \ref{sec:transfermat}, it can be checked that in the present case the operators $\mathbf{U}_1(\phi)$, $\mathbf{U}_2(\phi)$ take the form
\be 
\begin{split}
	\mathbf{U}_{\rm F} (\phi ) & = \mathbf{U}_1 (\phi)  \mathbf{U}_2 (\phi) , \\
    \mathbf{U}_1 (\phi) & = \prod_{m=1}^L \exp \left( -\ii \tau \mathbf{e}_{2m-1}  \right) = \prod_{m=1}^L \left( \mathbbm{1} + \frac{\exp (-\ii \sqrt{Q} \tau) -1}{\sqrt{Q} } \mathbf{e}_{2m-1} \right) \\
    & = \prod_{m=1}^L  \left( \frac{\exp (-\ii \sqrt{Q} \tau)+Q-1}{Q} + \frac{\exp (-\ii \sqrt{Q} \tau)-1}{Q} \sum_{a=1}^{Q-1} \mathbf{X}_m^a \right) , \\ 
    \mathbf{U}_2 (\phi) & = \prod_{m=1}^L \exp \left( -\ii \tau \mathbf{e}_{2m}  \right) = \prod_{m=1}^L  \left( \mathbbm{1} + \frac{\exp (-\ii \sqrt{Q} \tau) -1}{\sqrt{Q} } \mathbf{e}_{2m} \right) \\
    & =\prod_{m=1}^L \left( \frac{\exp (-\ii \sqrt{Q} \tau)+Q-1}{Q} + \frac{\exp (-\ii \sqrt{Q} \tau)-1}{Q} \sum_{a=1}^{Q-1} \left( \mathbf{Z}_m^\dag \mathbf{Z}_{m+1} \right)^a \right)  ,
\end{split}
\label{eq:Pottscircuits1}
\ee
where the spectral parameter $\phi$ is related to the ``period'' $\tau$ by
\be 
  \exp ( - \ii \sqrt{Q} \tau ) = 1 + \sqrt{Q} \frac{\sinh ( \eta \phi / \pi ) }{\sinh \big( \eta (\pi - \phi) / \pi \big)} .
  \label{eq:tautophi}
\ee 
Note in particular that the Floquet evolution operator $\mathbf{U}_{\rm F}(\phi)$ is of the same form as given in eq. \eqref{eq:ZQFloquet}. It is uniquely defined by the value of $\phi$ modulo arbitrary shifts by $\frac{2i\pi}{\eta}$, or by the value of $\tau$ modulo arbitrary shifts by $\frac{2\pi}{\sqrt{Q}}$. Furthermore, because of the hermiticity of the generators $\mathbf{e}_m$, the dynamics is unitary whenever $\tau \in \mathbb{R}$,
\be 
	\mathbf{U}_{\rm F} (\phi ) \mathbf{U}_{\rm F}^\dag (\phi ) = \mathbbm{1} , \quad \tau \in \mathbb{R} ,
\ee 
or equivalently the parameter $\phi$ must satisfy the following identity
\be 
	\left| 1 + \sqrt{Q} \frac{\sinh ( \eta \phi / \pi ) }{\sinh \big( \eta (\pi - \phi) / \pi \big)} \right| = 1 .
 \label{eq:unitarityphi}
\ee
The values of $\phi$ solving \eqref{eq:unitarityphi} are generally complex. However, for $Q=2$ or $Q=3$, some real solutions of are of particular interest as they connect to known models.

For $Q=2$, nontrivial real solutions to \eqref{eq:unitarityphi} are found as $\phi=\pm 2\pi$, corresponding to $\tau=\frac{\pi}{\sqrt{2}}$.
In this case, the evolution operator $\mathbf{U}_{\rm F}(\phi)$ commutes with the Hamiltonian 
\be 
\mathbf{H} = \sum_{j=1}^{2L} \mathbf{e}_{j} \,,
\ee 
which coincides with the Hamiltonian of the spin-1/2 XX model up to unitary transformation. 

Similarly, for $Q = 3$ nontrivial real solutions of \eqref{eq:unitarityphi} are found as $\phi=\pm 3\pi$, corresponding to $\tau=\frac{\pi}{\sqrt{3}}$. At this value of $\tau$, the circuit dynamics can be related to the Zamolodchikov--Fateev 19-vertex model \cite{ZF_1980}, as will be discussed in Sec. \ref{subsec:Potts19v}.

\subsection{Local conserved charges}

We follow the way of Section \ref{sec:transfermat} to construct two sets of local charges commuting with the circuit dynamics, $\mathbf{I}_{m,0}$ and $\mathbf{I}_{0,n}$, $m,n\in\mathbb{Z}_{>0}$. 
Using the normalisation of the R matrix \eqref{eq:Rmatnormalisation}, we can express the first two charges as
\be 
    \mathbf{I}_{1,0} = \frac{1}{f(\phi)} \sum_{j=1}^L \partial_{\lambda} \mathbf{R}_{j,j+1} (\lambda , 0, \phi) \mathbf{P}_{j,j+1} , \quad \mathbf{I}_{0,1} = \frac{1}{f(\phi)} \sum_{j=1}^L \partial_{\mu} \mathbf{R}_{j,j+1} (0, \mu , \phi) \mathbf{P}_{j,j+1} .
    \label{eq:densitycharges}
\ee 
We find (see Appendix \ref{app:Pottslocaldensity} for details)
\be 
    \mathbf{I}_{1,0} + \mathbf{I}_{0,1}= \frac{2}{Q} \big( \mathbf{Q}_1 +c_1 \big) ,
    \label{eq:Q_1}
\ee 

and 
\be 
    \mathbf{I}_{1,0} - \mathbf{I}_{0,1}  = \frac{2}{Q} \big( \mathbf{Q}^\prime_1 + c_2\big) ,
    \label{eq:Q_1prime}
\ee 
where $c_1$, $c_2$ are constant and
\be 
	\mathbf{Q}_1 = \sum_{j=1}^{2L} \mathbf{e}_{j} + \frac{\ii}{2\sqrt{Q}} \sin (\sqrt{Q} \tau) (-1)^{j} \left[ \mathbf{e}_{j} , \mathbf{e}_{j+1} \right]  - \frac{1}{\sqrt{Q}}\sin^2 \big( \frac{\sqrt{Q} \tau}{2} \big) \left\{ \mathbf{e}_{j} , \mathbf{e}_{j+1} \right\} ,  
	\label{eq:chargesTL}
\ee 
and 
\be 
   \mathbf{Q}^\prime_1 = \sum_{j=1}^{2L} -\frac{\ii \sin (\sqrt{Q} \tau ) }{2 \sqrt{Q}} \left[ \mathbf{e}_j , \mathbf{e}_{j+1} \right] + \frac{ (-1)^j \sin^2 (\sqrt{Q} \tau /2) }{\sqrt{Q}} \left\{ \mathbf{e}_j , \mathbf{e}_{j+1} \right\} \,.
\ee
In \cite{Gritsev_2022}, a set of conserved charges $\mathbf{Q}_1$, $\mathbf{Q}_2$, $\mathbf{Q}_3$ commuting with the dynamics \eqref{eq:Pottscircuits1} was constructed in terms of the generators $\mathbf{e}_j$, by explicitly computing the commutation with the evolution operator $\mathbf{U}_F$. Explicit expressions were given for $\mathbf{Q}_1$ and $\mathbf{Q}_2$, while the expression of $\mathbf{Q}_3$ is more involved. It is easy to check that our charge $\mathbf{Q}_1$ given by \eqref{eq:chargesTL} coincides with the one given in \cite{Gritsev_2022}. Furthermore, we check that the charge $(\mathbf{I}_{2,0} - \mathbf{I}_{0,2})$ coincides with the charge $\mathbf{Q}_2$ of \cite{Gritsev_2022}, up to a proportionality factor and constant. We believe that, similarly, we could recover the charge $\mathbf{Q}_3$ of \cite{Gritsev_2022}.  
Therefore, our construction recovers and extends the family of charges $\mathbf{Q}_m$ proposed in \cite{Gritsev_2022}, together with an additional family $\mathbf{Q}'_m$, given by the linear combination of the charges $\mathbf{I}_{m,0}$ and $\mathbf{I}_{0,m}$.

We now comment on the possibility to relate the discrete time evolution operator $\mathbf{U}_{\rm F}(\tau)$ to some quantum Hamiltonian acting in continuous time. 
In Floquet systems, what is commonly defined as the Floquet Hamiltonian $\mathbf{H}_{\rm F}$ is defined formally as $\mathbf{U}_{\rm F}(\tau) = \exp(i\tau \mathbf{H}_{\rm F})$, and is not local. In contrast, one can define a local Hamiltonian by taking the $\tau\to 0$ (Trotter) limit: this is nothing but the quantum Potts Hamiltonian $\mathbf{H}_1+\mathbf{H}_2$, which however does not commute with  $\mathbf{U}_{\rm F}(\tau)$ for generic $\tau$. 
A third possibility is to use the charges $\mathbf{Q}_j$, $\mathbf{Q}'_j$ defined above. Those are local operators (namely, sums of local densities), they are  furthermore hermitian, and by construction they commute with $\mathbf{U}_{\rm F}(\tau)$. They can therefore be considered as Hamiltonians generating some continuous time dynamics, sharing the same integrals of motion as the Floquet dynamics generated by $\mathbf{U}_{\mathrm{F}}(\tau)$.
Let us point that, in the $\tau\to 0$, all the Hamiltonians defined above (more precisely, the Floquet Hamiltonian, the Potts Hamiltonian and the sum $\mathbf{Q}_1+\mathbf{Q}_1'$) become proportional to each other.

\subsection{3-state Potts case and 19-vertex model}
\label{subsec:Potts19v}

We now come back to the connection mentioned at the end of Section \ref{sec:Potts/circuit}, between the 3-state Potts circuit with $\phi = 3\pi$ and the Zamolodchikov--Fateev 19-vertex model at root of unity $q = \exp \left( \frac{\ii \pi}{3} \right)$ \cite{ZF_1980, Vernier_2019}. 

The Zamolodchikov--Fateev 19-vertex model \cite{ZF_1980, Piroli_2016} can be obtained via transfer matrix fusion of the 6-vertex model \cite{kulish1981yang}. One of the conserved quantities (obtained via the logarithmic derivative of the transfer matrix) is a spin-1 Hamiltonian, which can be considered as the integrable spin-1 generalisation of the spin-$1/2$ XXZ model. As in the spin-$1/2$ case the model is defined in terms of a complex parameter $q$ relating to the underlying quantum group $U_{q}(sl_2)$. At the ``root of unity'' points $q^N = \pm 1$ it is conjectured to have a hidden Onsager algebra symmetry  \cite{YM_spectrum_XXZ, YM_Onsager}, which can be shown explicitly for $q=\exp \left( \frac{\ii \pi}{3} \right)$  \cite{Vernier_2019, YM_Onsager}.

Interestingly, the conserved quantities obtained from two-parameter transfer matrix \eqref{eq:TQP} consist of a subset of the generators of the Onsager algebra (up to a unitary transformation), which is not obvious at first sight.

To begin with, let us consider the following unitary transformation carried out by the operator
\be 
  \mathcal{U}^{(3)}_m = \frac{1}{\sqrt{3}} \bp 1 & 1 & 1 \\ 1 & \omega & \omega^2 \\  1 & \omega^2 & \omega \ep_m , 
\ee
with the third root of unity $\omega = \exp (2 \pi \ii /3)$.
The operator $\mathcal{U}^{(3)}_m$ transfers the 3-state Potts spin as follows,
\be 
\mathcal{U}^{(3)}_m \mathbf{X}_m {\mathcal{U}^{(3)}_m}^\dag = \mathbf{Z}_m^\dag ,  \quad \mathcal{U}^{(3)}_m \mathbf{Z}_m {\mathcal{U}^{(3)}_m}^\dag = \mathbf{X}_m .
\ee

In addition, we need another unitary operator
\be 
    \mathcal{V}_m = \bp 0 & 1 & 0 \\ 1 & 0 & 0 \\ 0 & 0 & 1 \ep = \mathcal{V}^\dag_m , \quad  \mathcal{V}_m^2 = \mathbbm{1}_m . 
\ee

The 19-vertex model R matrix with $q = \exp \left( \frac{\ii \pi}{3} \right)$ is obtained as a special case of the two-parameter R matrix with $\phi = 3\pi$ depicted in Fig. \ref{fig:R_mat} after the unitary transformation,
\be
    \tilde{\mathbf{R}}_{a,b} (\lambda , \mu ) = - \mathcal{V}_a \mathcal{V}_b \mathcal{U}^{(3)}_a \mathcal{U}^{(3)}_b \mathbf{R}_{a,b} (\lambda , \mu , \phi = 3\pi) {\mathcal{U}^{(3)}_a}^\dag {\mathcal{U}^{(3)}_b}^\dag \mathcal{V}_a \mathcal{V}_b .
\ee

When $\mu= \lambda$, we recover the renowned 19-vertex R matrix at root of unity $q = \exp (\ii \pi /3)$  \cite{ZF_1980}, 
\be
\resizebox{.92\linewidth}{!}{
$
\tilde{\mathbf{R}}_{a,b} ( - \lambda , - \lambda ) = \bp a(\lambda) & 0 & 0 & 0 & 0 & 0 & 0 & 0 & 0 \\ 0 & b(\lambda) & 0 & c(\lambda) & 0 & 0 & 0 & 0 & 0 \\ 0 & 0 & d(\lambda) & 0 & e(\lambda) & 0 & g & 0 & 0 \\ 0 & c(\lambda) & 0 & b(\lambda) & 0 & 0 & 0 & 0 & 0 \\  0 & 0 & e(\lambda) & 0 & f(\lambda) & 0 & e(\lambda) & 0 & 0 \\ 0 & 0 & 0 & 0 & 0 & b(\lambda) & 0 & c(\lambda) & 0 \\ 0 & 0 & g & 0 & e(\lambda) & 0 & d(\lambda) & 0 & 0 \\ 0 & 0 & 0 & 0 & 0 & c(\lambda) & 0 & b(\lambda) & 0 \\  0 & 0 & 0 & 0 & 0 & 0 & 0 & 0  & a (\lambda) \ep = \mathcal{R} (\lambda) , 
$}
\ee
where the coefficients are defined
\be 
\begin{split}
    & a (\lambda) = [u+1][u+2] = \frac{1}{3} \left( 1 + 2 \cos \frac{2\lambda}{3} \right) , \\ 
    & b (\lambda) = [u][u+1] = \frac{1}{3} \left( 1 - \cos \frac{2\lambda}{3} + \sqrt{3}\sin \frac{2\lambda}{3}  \right) , \\
    & c (\lambda) = \cos \frac{\lambda}{3} + \frac{1}{\sqrt{3}} \sin \frac{\lambda}{3} , \quad d (\lambda) = [u-1][u] = \frac{2}{3} \sin \frac{\lambda}{3} \left( \sin \frac{\lambda}{3} - \sqrt{3} \cos \frac{\lambda}{3} \right) , \\
    & e(\lambda) = [2] [u] = \frac{2}{\sqrt{3}} \sin \frac{\lambda}{3} , \quad f (\lambda) = b (\lambda) +[2] = \frac{1}{3} \left( 4 - \cos \frac{2\lambda}{3} + \sqrt{3}\sin \frac{2\lambda}{3}  \right) , \\
    & g = [2] = 1 , \quad u = \frac{\lambda}{\pi } ,
\end{split}
\ee 
with $q$-number defined as
\be 
[u] = \frac{q^u - q^{-u} }{q - q^{-1} } .
\ee 

Another intriguing fact is that the conserved quantities of the 19-vertex model at root of unity $q = \exp (\ii \pi /3)$ can be expressed in terms of the Temperley-Lieb algebra generators \cite{Fendley_2020}. To see this, we define the 19-vertex transfer matrix
\be
    \mathcal{T} (\lambda) = \mathrm{Tr}_a \left( \prod_{j=1}^L \mathcal{R}_{a,j} (\lambda ) \right) ,
\ee
and the first local conserved quantity (``the spin-1 ZF Hamiltonian'') becomes
\be
    \mathbf{H}^{\rm ZF} = \partial_\lambda \log \mathcal{T} (\lambda) = - \big( \mathbf{I}_{1,0} + \mathbf{I}_{0,1} \big) , 
\ee
due to the factorisation property of the two-parameter transfer matrix \eqref{eq:factorisation}.

\be 
    \mathbf{H}^{\rm ZF} = -\frac{2}{3} \mathcal{V} \mathcal{U}^{(3)}  \left[ \sum_{m=1}^{2L} \Big( \mathbf{e}_{m} - \frac{1}{\sqrt{3}} \{ \mathbf{e}_m , \mathbf{e}_{m+1} \} -\frac{1}{2\sqrt{3}} \Big) \right] {\mathcal{U}^{(3)}}^\dag  \mathcal{V} ,
\ee
where the unitary transformations are
\be 
\mathcal{U}^{(3)} = \prod_{m=1}^L \mathcal{U}^{(3)}_m , \quad \mathcal{V} = \prod_{m=1}^L \mathcal{V}_{m} .
\ee
The ZF Hamiltonian at root of unity $q = \exp (\ii \pi /3)$ can therefore be transformed into a special case of \eqref{eq:chargesTL} with $Q=3$ and $\tau = \pi / \sqrt{3}$. The Hamiltonian can be expressed in terms of spin-1 operators as well in a compact way, as shown in App. \ref{app:ZF}. 
More generally, the local charges $\mathbf{I}_{0,m}+\mathbf{I}_{m,0}$ generated by $\mathcal{T}(\lambda)$ recover the local conserved charges of the ZF spin 1 Hamiltonian derived from the usual spin-1 transfer matrix, while the charges $\mathbf{I}_{0,m}-\mathbf{I}_{m,0}$ form a mutually commuting subset of the Onsager symmetry generators. 
This connection is in fact part of a more general connection between solutions of the star-triangle equation and higher-spin descendants of the six-vertex model, which is currently under investigation.

\section{Example: $\mathbb{Z}_Q$ circuits}
\label{sec:AT}

Besides the $Q$-state Potts model, which possesses the $S_Q$ symmetry, there exist solutions to the star-triangle relation \eqref{eq:STR} with $\mathbb{Z}_Q$ symmetry \cite{AuYangPerk}. The most renowned one has been originally derived by Fateev and Zamolodchikov \cite{Fateev_1982_1, Fateev_1982_2, Albertini_1992_2}, and takes the form
\be 
\begin{split}
  & K_{\rm FZ}(\theta ; a , b) = 1 , \quad a-b=0 ,\\
  & K_{\rm FZ}(\theta ; a , b) = \prod_{m=0}^{|a-b| -1} \frac{\sin (\frac{\pi m}{Q} + \frac{\theta}{2Q} )}{\sin (\frac{\pi (m+1)}{Q} - \frac{\theta}{2Q} )} , \quad a-b \neq 0 .
\end{split}
\label{eq:STRZF}
\ee

For $Q = 3$, \eqref{eq:STRZF} coincides with \eqref{eq:STRPotts} up to normalisation factor. 
As pointed earlier, this is due to the fact that the $\mathbb{Z}_3$ symmetry together with the charge conjugation symmetry $K_{\rm FZ}(\theta ; a,b)= K_{\rm FZ}(\theta ; b,a)$ generate the symmetric group $S_3$, which is the symmetry of the 3-states Potts model.
In contrast, when $Q \geq 4$, \eqref{eq:STRZF} and \eqref{eq:STRPotts} become different. More specifically, \eqref{eq:STRZF} with $Q = 4$ is related to a critical Ashkin--Teller model \cite{Ashkin_Teller_1943, Kohmoto_1981, Rittenberg_1987, Alcaraz_1987, Alcaraz_1988}. When $Q=4$,
\be 
K_{\rm FZ4}(\theta ; a - b) = \begin{cases}
      1, & a-b=0 , \\
      \frac{\sin (\theta / 8 )}{\sin (\pi / 4- \theta/8)}, & |a-b| = 1 \, \mathrm{or} \, 3 , \\
      \frac{\tan (\theta / 8 )}{\tan (\pi / 4- \theta/8)}, & |a-b| = 2 .
    \end{cases}
\label{eq:STRZF4}
\ee 
The critical Ashkin-Teller Hamiltonian is obtained by considering the first local conserved charge in the limit $\phi \to 0$, which is shown in Appendix \ref{app:criticalAT}.

We focus on the $\mathbb{Z}_4$ circuit now. Similar to the Potts case, the $\mathbb{Z}_4$ circuit is built on the Floquet evolution operator such that
\be 
	\mathbf{U}_{\rm F} (\phi) = \mathbf{U}_1 (\phi) \mathbf{U}_2 (\phi) ,
\ee
which is closely related to the two-parameter transfer matrix such that
\be 
\begin{split}
	& \mathbf{T} (0,\phi , \phi ) = \mathbf{V} (\phi , \phi ) \mathbf{W} (0,\phi ) , \\
	& \mathbf{V} (\phi,\phi ) = \mathbf{G}^{-1}  \mathbf{U}_1 (\phi )  = \mathbf{U}_1 (\phi )  \mathbf{G}^{-1} =  \mathbf{G}^{-1} \prod_{m=1}^L \mathbf{v}_m  \\
	& \mathbf{W} (0, \phi) = \mathbf{U}_2 (\phi ) = \prod_{m=1}^L \mathbf{w}_{m , m+1} ,
\end{split}
\label{eq:ATcircuits1}
\ee
where the local quantum gates are
\be 
    \mathbf{v}_m = \mathbbm{1} + K_{\rm FZ4} (\phi ; 1) \big( \mathbf{X}_m + \mathbf{X}_m^\dag \big) + K_{\rm FZ4} (\phi ; 2) \mathbf{X}_m^2 ,
\ee
\be 
\begin{split}
    \mathbf{w}_{m,m+1} = & \frac{1}{4} \big(1+ 2 K_{\rm FZ4} (\pi-\phi ; 1) + K_{\rm FZ4} (\pi-\phi ; 2) \big) +\\
    & \frac{1}{4} \big( 1-K_{\rm FZ4} ( \pi-\phi ; 2) \big) \big( \mathbf{Z}_m^\dag \mathbf{Z}_{m+1} + \mathbf{Z}_m \mathbf{Z}_{m+1}^\dag \big) \\
    & + \frac{1}{4} \big(1 - 2 K_{\rm FZ4} (\pi-\phi ; 1) + K_{\rm FZ4} (\pi-\phi ; 2) \big) \mathbf{Z}^2_m \mathbf{Z}^2_{m+1} .
\end{split}
\ee

The evolution operators $\mathbf{U}_1(\phi)$ and $\mathbf{U}_2(\phi)$ are of the generic form \eqref{eq:ZQFloquet}. However, unlike the the Potts case \eqref{eq:Pottscircuits1}, where there exist sets of $\phi$ as solutions to \eqref{eq:tautophi} that guarantee the quantum circuits to be unitary, there is no $\phi$ that makes the quantum circuits \eqref{eq:ATcircuits1} unitary, except for the trivial cases when $\phi = 8 n \pi$ or $\phi = 4 \pi + 8 n \pi$ after rescaling.

Even though the integrable quantum circuits obtained using the Fateev--Zamolodchikov star-triangle relation are not unitary in general, the integrability has not been shown in previous literature up to our knowledge, which could potentially be intriguing to study the physical properties. Similar non-unitary integrable quantum circuits have been studied in \cite{YM_Floquet, Yashin_2022}, closely related to the non-unitary conformal field theory. It would be interesting to see if the $\mathbb{Z}_Q$ circuits can be understood analogously, which we will not discuss in details here. Sometimes the non-unitary integrable quantum circuits are also the completely positive trace-preserving (CPTP) maps \cite{Prosen_2021}, which has a closely relation to the open quantum systems. It will be useful to investigate whether the non-unitary quantum circuits obtained from the star-triangle relations are the CPTP maps, which we intend to study later.

\section{Conclusion}
\label{sec:conclusion}

In this article we studied the integrable structure of quantum circuits in the form of Fig. \ref{fig:quantum_circuit_demo}, which can be considered as the Floquet dynamics of a time-dependent Potts-like quantum Hamiltonian. We used the renowned star-triangle relation to construct families of two-parameter transfer matrices that commute with the Floquet evolution operator, underlying the integrable structure. The quantum circuits are obtained by taking the spectral parameters of the two-parameter transfer matrix to special values. 

Compared to the known example of integrable quantum circuits of brick-wall type, whose construction is based on Yang-Baxter integrable vertex models \cite{Destri_1989, DDV, Vanicat, YM_Floquet}, the quantum circuits studied in this article indeed share a certain resemblance. However, even though we have shown that the two-parameter transfer matrices can be formulated as the row-to-row transfer matrices of certain vertex models in Sec. \ref{sec:transfermat}, the staggering of spectral parameters leading to a circuit geometry takes place in our construction between the internal parameters entering the definition of each R matrix, rather than between odd and even sites of the vertex model as in the case in the brick-wall approach \cite{Destri_1989, DDV, Vanicat, YM_Floquet}. 
This difference is what makes our construction new, and allows for a systematic construction of new families of integrable quantum circuits based on solutions to the star-triangle relations.

We would like to comment in particular on the recent work by Bazhanov and Sergeev \cite{Bazhanov_2022}, where an alternative description of the six-vertex model was given, involving an underlying spin model satisfying the star-triangle relation \cite{Bazhanov_2022}. From there one might follow a similar approach to that of the present work, namely constructing an integrable circuit dynamics from an inhomogeneous two-row transfer matrix satisfying the star-triangle relation. Similar to the case reported in this article, we stress, however, that the corresponding circuits are {\it not} equivalent to those constructed using a brick-wall ``trotterization'' of the six-vertex model \cite{Destri_1989, DDV, Vanicat, YM_Floquet}. The reason is again that the mapping described in \cite{Bazhanov_2022} assigns each site of the six-vertex model to a pair of sites of the underlying star-triangle model. The staggering of spectral parameters which we use in our construction therefore does not break the translation invariance of the underlying vertex model, moreover it needs to be fine-tuned in order to be compatible with a six-vertex formulation.

In this work we focused on two families of $Q$-states quantum circuits. The first is associated with the $Q$-states Potts model, for which we proved the conjectured integrability using the star-triangle relation of the Potts model \cite{Pokrovsky_1982}, and found an additional set of conserved charges expressed in terms of Temperley--Lieb generators. In the case of 3-state Potts, we presented a connection between the integrable quantum circuit and the integrable 19-vertex model \cite{ZF_1980}, which is part of a larger connection currently under investigation. The second family of circuits, dubbed $\mathbb{Z}_Q$ circuits, results from the Fateev--Zamolodchikov $\mathbb{Z}_Q$ solution of the star-triangle relation \cite{Fateev_1982_2}, and yields a different integrable quantum circuit that for $Q=4$ is closely related to the critical Ashkin-Teller spin chain.
Beyond these two examples, our construction should work for more general solutions of the star-triangle equation \cite{AuYangPerk}, and we leave the study of the corresponding circuits as an interesting perspective for future investigation.

There are still many aspects of the integrable quantum circuits in the form of Fig. \ref{fig:quantum_circuit_demo} that need to be investigated. The first natural question deals with the spectrum of the Floquet evolution operator, and whether it can be computed exactly using the toolbox of integrability. For the various kinds of circuits considered in this work, we expect that this can be achieved in a variety of ways. For the Potts circuits of Sec. \ref{sec:Potts}, one could try a similar approach to that of \cite{Albertini_1992_1}, where Bethe ansatz equations were obtained for the $Q=3$ Potts model. A a more indirect, but effective way is to use the representation of these models in terms of Temperley-Lieb algebra, amenable to a Bethe ansatz treatment due to the six-vertex representation of the latter \cite{Baxter_1982}. For the $\mathbb{Z}_Q$ circuits of Section \ref{sec:AT}, one should follow the approach of \cite{Albertini_1992_2}, where Bethe ansatz equations were obtained for $\mathbb{Z}_Q$ invariant models.

The next question would be studying the physical properties of quantum quenches in the circuits. The time evolution from certain initial product states could potentially be realised in recent experiments \cite{Lanyon_2011, Mi_2021} and the quantum integrability that we used can be a useful tool \cite{Caux_2013, Caux_2016}. In the case of vertex models, by using the boundary Yang-Baxter equations and a Wick rotation, the ``integrable quenches'' \cite{Piroli_2017} are investigated, and it is possible to obtain analytic results for the late-time steady states and various correlation functions for certain initial states. The same approach has been studied in the integrable brickwork quantum circuits setting recently \cite{VernierdGGE}. We anticipate that similar ``integrable quenches'' also exist in the star-triangle circuits using the boundary star-triangle relation \cite{Zhou_1997}.

Moreover, the field theory limit of the quantum circuits is also interesting, since the brick-wall quantum circuits are initially studied as the lattice regularisation of the field theories \cite{DDV, Reshetikhin_1994}. The generalisation of the brick-wall quantum circuits has been proposed in \cite{YM_Floquet}, while it is not clear how it can be extended to the quantum circuits considered in this article, cf. Fig. \ref{fig:quantum_circuit_demo}. All these questions remain to be studied and answered, which we intend to do in future works.

\section*{Acknowledgment}

The work of Y.M. was partially supported by World Premier International Research Center Initiative (WPI), MEXT, Japan. Y.M. also acknowledges the support from the GGI BOOST fellowship. Y.M. is grateful to Vladimir Gritsev and Denis Kurlov for collaborations on related topics. We would like to thank Filippo Colomo, Paul Fendley, Jesper Jacobsen, Jules Lamers, Hosho Katsura, Vincent Pasquier for useful discussions, and especially Balázs Pozsgay for early collaboration on closely related topics.


\begin{appendix}
\section{Diagrammatic derivations of some formulae}
\label{app:diagramma}

\subsection{Diagrammatic derivation of the Yang--Baxter relation}
\label{app:YBE}

\begin{figure}[H]
\centering
\includegraphics[width=.9\linewidth]{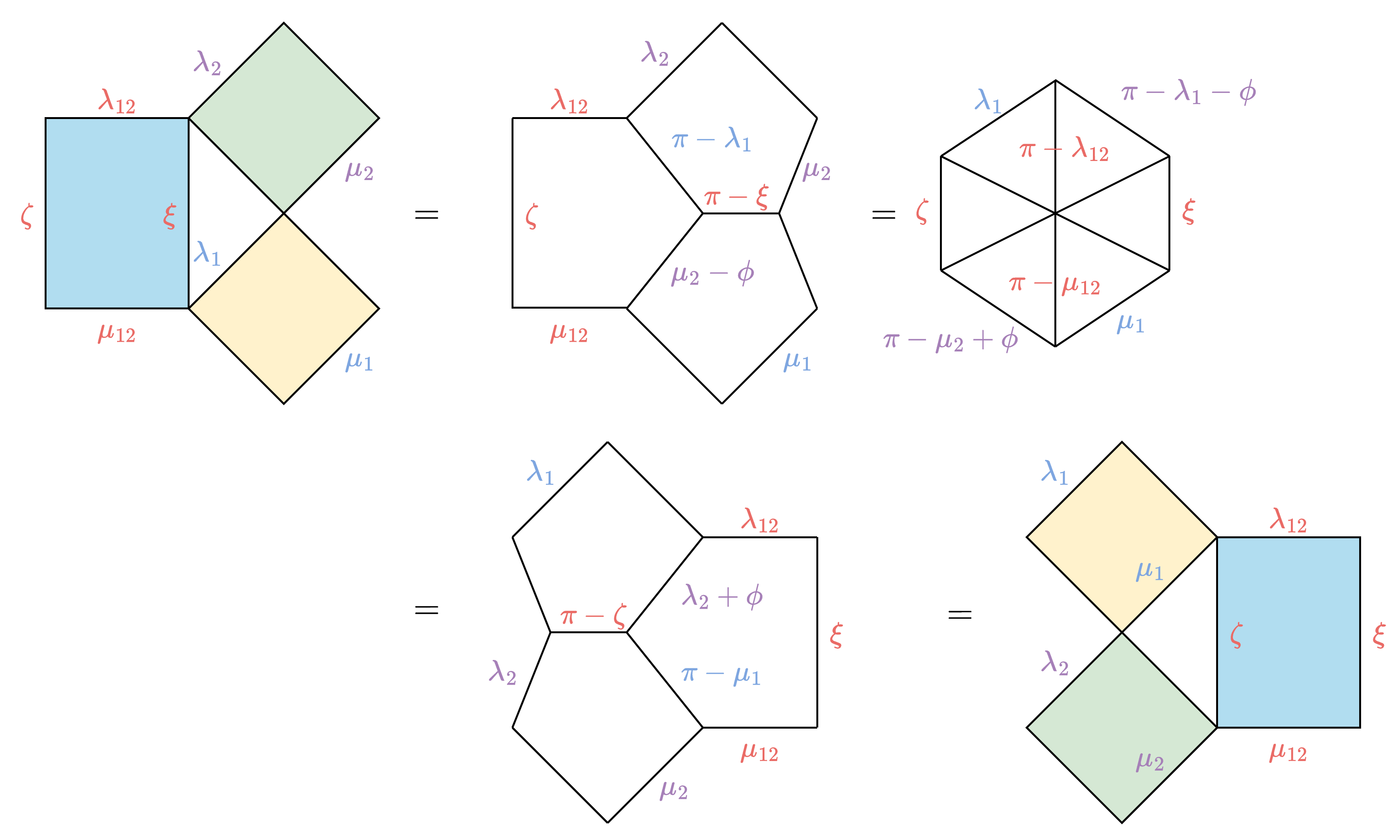}
\caption{The proof of the Yang--Baxter relation \eqref{eq:YBE1} by recursively applying the star-triangle relation \eqref{eq:STR}. }
\label{fig:YBEproof}
\end{figure}

The Yang--Baxter relation of the R matrix \eqref{eq:Rmatdef} is proven directly from the star-triangle relation \eqref{eq:STR}. By first applying the star-triangle relation in the white triangle in between of the coloured rectangular, the detailed derivation is summarised in Fig. \ref{fig:YBEproof}.

\subsection{Diagrammatic derivation of the self-dual relation}
\label{app:selfdual}

\begin{figure}[H]
\centering
\includegraphics[width=.95\linewidth]{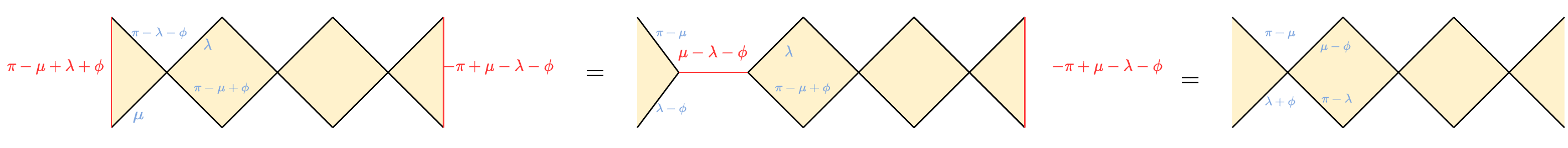}
\caption{The proof of the ``self-dual'' property of the two-parameter transfer matrix \eqref{eq:selfdual}. }
\label{fig:selfdual}
\end{figure}

We start with inserting an identity operator, which is decomposed into two parts, the red line on the left and its inverse on the right, since we are assuming the periodic boundary condition here. By pushing the operator to the left using the star-triangle relation \eqref{eq:STR}, the spectral parameters of the R matrix change accordingly, cf. Fig. \ref{fig:selfdual}. Eventually, the operator cancels with its inverse on the right end, changing the spectral parameters of the transfer matrix, i.e. the self-dual relation in \eqref{eq:selfdual}.

\subsection{Diagrammatic derivation of Eq.~\eqref{eq:factor1} }
\label{app:factorder}

\begin{figure}[H]
\centering
\includegraphics[width=.8\linewidth]{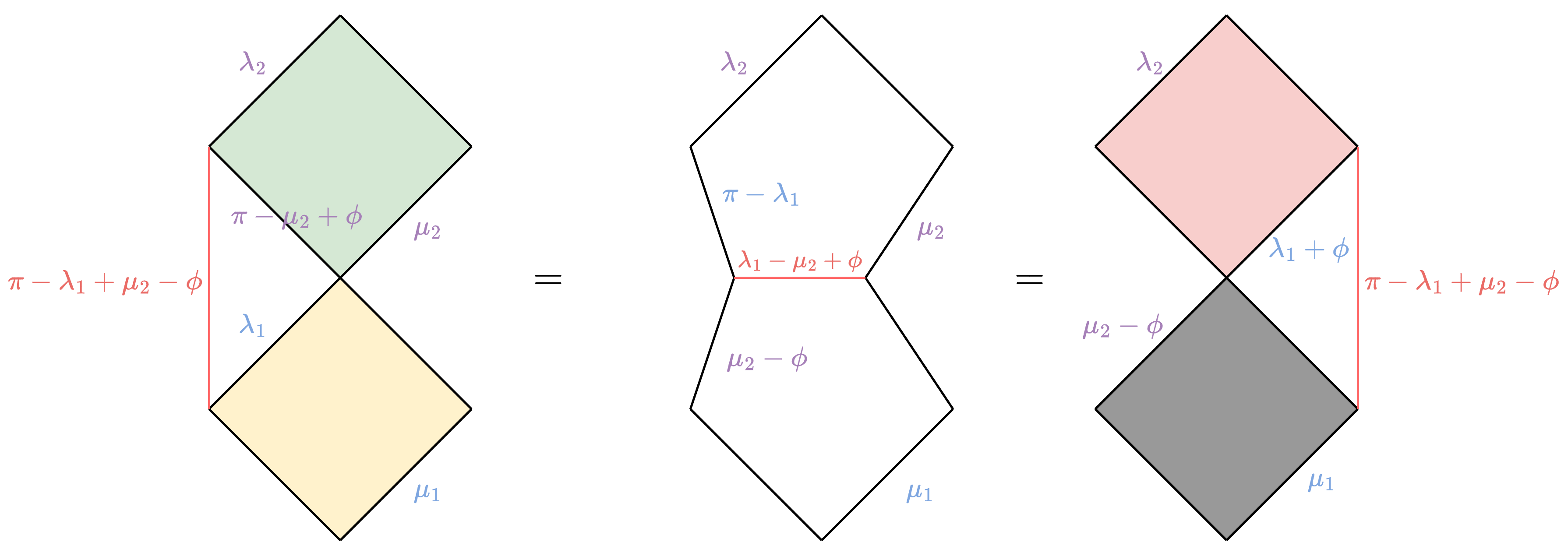}
\caption{The proof of \eqref{eq:factor1} in terms of diagrams. }
\label{fig:factor1}
\end{figure}

\section{Explicit form of spin-1 ZF Hamiltonian at root of unity}
\label{app:ZF}

We follow the example of \cite{Vernier_2019} and use the spin-1 $\mathfrak{sl}_2$ operators 
\be 
    \mathcal{S}^+_m = \bp 0 & 1 & 0 \\ 0 & 0 & 1 \\ 0 & 0 & 0 \ep_m , \quad \mathcal{S}^-_m = \bp 0 & 0 & 0 \\ 1 & 0 & 0 \\ 0 & 1 & 0 \ep_m 
\ee
to rewrite the spin-1 ZF Hamiltonian at root of unity $q = \exp (\ii \pi /3)$, i.e.
\be 
    \mathbf{H}^{\rm ZF} = \frac{2}{3\sqrt{3}} \sum_{m=1}^L \sum_{a=1}^2 \left[ (-1)^a (\mathcal{S}^+_m \mathcal{S}^-_{m+1})^a +(-1)^a (\mathcal{S}^-_m \mathcal{S}^+_{m+1})^a + \frac{1}{3(1+\omega^{-a})} \mathbf{Z}_m^a + \frac{1}{3} \right] ,
\ee
where $\mathbf{Z}_m$ are the 3-state Potts operator in \eqref{XmZmdef} and $\omega = \exp (2\ii \pi /3)$ is the third root of unity.

\section{Local density of charges in $Q$-state Potts circuits}
\label{app:Pottslocaldensity}

By directly calculating the local charge densities and expressing them in terms of the affine TL generators, the local charge densities \eqref{eq:densitycharges} become
\be 
\begin{split}
    \frac{1}{f(\phi)}  \partial_{\lambda} \mathbf{R}_{j,j+1} (\lambda , 0, \phi) \mathbf{P}_{j,j+1} & = \frac{1}{Q} \left[ \mathbf{e}_{2j-1} + \mathbf{e}_{2j} + \frac{2\ii \sin (\sqrt{Q} \tau ) }{2 \sqrt{Q}} \left[ \mathbf{e}_{2j-1} , \mathbf{e}_{2j} \right] \right. \\
    & \left. - \frac{2\sin^2 (\sqrt{Q} \tau /2) }{\sqrt{Q}} \left\{ \mathbf{e}_{2j-1} , \mathbf{e}_{2j} \right\} -\frac{2 + e^{-\ii \sqrt{Q} \tau } }{\sqrt{Q}} \right],
\end{split}
\ee 
and 
\be 
\begin{split}
    \frac{1}{f(\phi)} \partial_{\mu} \mathbf{R}_{j,j+1} (0 , \mu, \phi) \mathbf{P}_{j,j+1} & = \frac{1}{Q} \left[ \mathbf{e}_{2j} + \mathbf{e}_{2j+1} + \frac{2\ii \sin (\sqrt{Q} \tau ) }{2 \sqrt{Q}} \left[ \mathbf{e}_{2j} , \mathbf{e}_{2j+1} \right] \right. \\
    & \left. - \frac{2\sin^2 (\sqrt{Q} \tau /2) }{\sqrt{Q}} \left\{ \mathbf{e}_{2j} , \mathbf{e}_{2j+1} \right\} -\frac{2 + e^{\ii \sqrt{Q} \tau } }{\sqrt{Q}} \right] ,
\end{split}
\ee 
where we have used the relation between $\phi$ and $\tau$ \eqref{eq:tautophi}.

By summing up the local density, and telescoping the sum, we arrive at
\be 
\begin{split}
    \mathbf{I}_{1,0} + \mathbf{I}_{0,1} = \frac{2}{Q} & \Bigg[ \Big( \sum_{j=1}^{2L} \mathbf{e}_j + (-1)^j \frac{\ii \sin (\sqrt{Q} \tau ) }{2 \sqrt{Q}} \left[ \mathbf{e}_j , \mathbf{e}_{j+1} \right] \\
    &  - \frac{\sin^2 (\sqrt{Q} \tau /2) }{\sqrt{Q}} \left\{ \mathbf{e}_j , \mathbf{e}_{j+1} \right\} \Big) - \frac{2-\cos(\sqrt{Q}\tau)}{\sqrt{Q}}L \Bigg] ,
\end{split}
\ee 
and 
\be 
\begin{split}
    \mathbf{I}_{1,0} - \mathbf{I}_{0,1} = \frac{2}{Q} & \Bigg[ \Big(  \sum_{j=1}^{2L} -\frac{\ii \sin (\sqrt{Q} \tau ) }{2 \sqrt{Q}} \left[ \mathbf{e}_j , \mathbf{e}_{j+1} \right]  \\
    &   + \frac{ (-1)^j \sin^2 (\sqrt{Q} \tau /2) }{\sqrt{Q}} \left\{ \mathbf{e}_j , \mathbf{e}_{j+1}\right\} \Big) + \frac{\ii \sin (\sqrt{Q}\tau)}{\sqrt{Q}} L  \Bigg].
\end{split}
\ee 
The two constants in \eqref{eq:Q_1} and \eqref{eq:Q_1prime} thus are
\be 
    c_1 = - \frac{2-\cos(\sqrt{Q}\tau)}{\sqrt{Q}} L , \quad c_2 = \frac{\ii \sin (\sqrt{Q}\tau)}{\sqrt{Q}} L .
\ee

\section{Explicit form of critical Ashkin-Teller model}
\label{app:criticalAT}

In the limit $\phi \to 0$, the two sets of local charges from the two-parameter transfer matrix coincide due to the self-duality \eqref{eq:selfdual},
\be 
    \phi = 0 \quad \Rightarrow \quad \mathbf{I}_{m,0} = \mathbf{I}_{0,m} .
\ee

In order to compare with the Ashkin-Teller Hamiltonian in the literature \cite{Alcaraz_1987, Rittenberg_1987}, we introduce the unitary transformation
\be 
\begin{split}
  & \mathcal{U}^{(4)}_m = \frac{1}{2} \bp 1 & 1 & 1 & 1 \\ 1 & \ii & -1 & -\ii \\  1 & -1 & 1 & -1 \\ 1 & - \ii & -1 & \ii \ep_m , \\
    & \mathcal{U}^{(4)} = \prod_{m=1}^L \mathcal{U}^{(4)}_m .
\end{split}
\ee
Therefore, the critical Ashkin-Teller Hamiltonian becomes
\be 
\begin{split}
    \mathbf{H}^{\rm AT} = & 4\sqrt{2} \mathcal{U}^{(4)} \mathbf{I}_{1,0} {\mathcal{U}^{(4)}}^\dag = \sum_{m=1}^L \left[ \mathbf{Z}_m + \mathbf{Z}_m^\dag + \frac{1}{\sqrt{2}} \mathbf{Z}_m^2 \right. \\ 
    & \left. + \mathbf{X}_m^\dag \mathbf{X}_{m+1} + \mathbf{X}_m \mathbf{X}_{m+1}^\dag + \frac{1}{\sqrt{2}} \mathbf{X}_m^2 \mathbf{X}_{m+1}^2 - \big(2+\frac{1}{\sqrt{2}} \big)  \right] ,
\end{split}
\label{eq:ATHamiltonian}
\ee
where $\mathbf{Z}_m$ and $\mathbf{X}_m$ are 4-state Potts operators in \eqref{XmZmdef}.

The Ashkin-Teller Hamiltonian obtained here \eqref{eq:ATHamiltonian} belongs to only one point of the self-dual critical line of the phase diagram \cite{Rittenberg_1987}. In addition, the Hamiltonian might appear in different guises in the literature. For instance, it is also possible to express the Hamiltonian \eqref{eq:ATHamiltonian} as a spin-1/2 ladder \cite{Kohmoto_1993, Kohmoto_1994}. A non-Hermitian version of the Ashkin-Teller model has been shown to be equivalent to the dissipative quantum Ising chain \cite{Katsura_2019}.

\end{appendix}

\bibliographystyle{quantum}
\bibliography{IntSTR}

\end{document}